\documentclass{aa}  
\usepackage{orcidlink}
\usepackage{graphicx}
\usepackage{txfonts}
\usepackage{lipsum}
\usepackage{subcaption}         
\usepackage{lscape}             
\usepackage{placeins}           
\usepackage{tablefootnote}
\usepackage[switch, modulo]{lineno}
\renewcommand{\linenumbers}{}                            

\begin{document}

   \title{Characterizing Clumpy Galaxies in Structures at Cosmic Noon in the Abell 2744 Field}

   \author{Emmet Golden-Marx\orcidlink{0000-0001-5160-6713}\thanks{\email{emmet.goldenmarx@inaf.it}}\inst{\ref{INAF-OAPd}}
    \and Benedetta Vulcani\orcidlink{0000-0003-0980-1499}\inst{\ref{INAF-OAPd}}
    \and Marco Gullieuszik\orcidlink{0000-0002-7296-9780}\inst{\ref{INAF-OAPd}}
    \and Giorgia Di Rosa\orcidlink{0009-0001-9416-0923}\inst{\ref{Ferrara},\ref{INFN},\ref{INAF-OASS}}
   \and Peter J. Watson\orcidlink{0000-0003-3108-0624}\inst{\ref{INAF-OAPd},\ref{OCA-NICE}}
   \and Finn Giddings\orcidlink{0009-0003-2158-1246}\inst{\ref{UH}}
   \and Lu Shen\orcidlink{0000-0001-9495-7759}\inst{\ref{TexasAM},\ref{Fudan}}
   \and Carmelo Arcidiacono\orcidlink{0000-0003-0142-8108}\inst{\ref{INAF-OAPd}}
   \and Lorenzo Bazzanini\orcidlink{0000-0003-0727-0137}\inst{\ref{INAF-IRA},\ref{Ferrara}}
   \and Pietro Bergamini\orcidlink{0000-0003-1383-9414}\inst{\ref{INAF-OASS}}
   \and Elisa Bortolas\orcidlink{0000-0001-9458-821X}\inst{\ref{INAF-OAPd}}
   \and Laura Bisigello\orcidlink{0000-0003-0492-4924}\inst{\ref{INAF-OAPd}}
   \and Ben Forrest\orcidlink{0000-0001-6003-0541}\inst{\ref{CUBoulder},\ref{UCDavis}}
   \and Giorgia Girardi\orcidlink{0009-0005-6156-4066}\inst{\ref{UniPd},\ref{INAF-OAPd}}
   \and Andrea Grazian\orcidlink{0000-0002-5688-0663}\inst{\ref{INAF-OAPd}}
   \and Weida Hu\orcidlink{0000-0003-3424-3230}\inst{\ref{SHAO}}
   \and Brian C. Lemaux\orcidlink{0000-0002-1428-7036}\inst{\ref{Gemini},\ref{UCDavis}}
   \and Marco Lombardi\orcidlink{0000-0002-3336-4965}\inst{\ref{Milano}}
   \and Elisa Portaluri\orcidlink{0000-0002-1242-7242}\inst{\ref{INAFOAA}}
   \and Piero Rosati\orcidlink{0000-0002-6813-0632}\inst{\ref{Ferrara},\ref{INAF-OASS}}
   \and Derek Sikorski\orcidlink{0000-0001-5796-2807}\inst{\ref{UH}}
   \and Matteo Simioni\orcidlink{0000-0002-8830-1259}\inst{\ref{INAF-OAPd}}}

    \institute{
    INAF - Osservatorio Astronomico di Padova, Vicolo Osservatorio 5, 35122 Padova, Italy\label{INAF-OAPd}
   \and
   Dipartimento di Fisica e Scienze della Terra, Universit\`a degli Studi di Ferrara, Via Giuseppe Saragat 1, 44122 Ferrara, Italy \label{Ferrara}
   \and
   INFN - Sezione di Ferrara, via Saragat 1, I-44122, Ferrara, Italy\label{INFN}
   \and
   INAF -- Osservatorio di Astrofisica e Scienza dello Spazio di Bologna, Via Piero Gobetti 93/3, 40129 Bologna, Italy \label{INAF-OASS}   
   \and
    Universit\'e C\^ote d'Azur, Observatoire de la Côte d’Azur, CNRS, Laboratoire Lagrange, 06000 Nice, France\label{OCA-NICE}
    \and
    Institute for Astronomy, University of Hawai‘i, 2680 Woodlawn Drive, Honolulu, HI 96822, USA\label{UH}
   \and 
   Department of Physics and Astronomy, Texas A\&M University, College Station, TX 77843, USA\label{TexasAM}
   \and 
   Fudan University, Shanghai 200433, People's Republic of China \label{Fudan}
   \and
   INAF-IRA -- Istituto di Radioastronomia, via Gobetti 101, I-40129, Bologna, Italy \label{INAF-IRA}
   \and
   Department of Astrophysical \& Planetary Sciences, University of Colorado, Boulder, 2000 Colorado Ave, Boulder, CO 80309, USA\label{CUBoulder}  
   \and
   Department of Physics and Astronomy, University of California, Davis, One Shields Avenue, Davis, CA 95616, USA \label{UCDavis}
   \and 
   Dipartimento di Fisica e Astronomia 'Galileo Galilei', Università degli Studi di Padova, Italy\label{UniPd}
   \and
   Key Laboratory for Research in Galaxies and Cosmology, Shanghai Astronomical Observatory, Chinese Academy of Sciences, 80 Nandan Road, Shanghai 200030, People’s Republic of China\label{SHAO}
   \and
   Gemini Observatory, NSF NOIRLab, 670 N. A'ohoku Place, Hilo, HI 96720, USA\label{Gemini}
   \and 
   Università degli Studi di Milano, Dipartimento di Fisica, via Celoria 16, I-20133 Milano, Italy\label{Milano}
   \and
   INAF - Osservatorio Astronomico d'Abruzzo, via M. Maggini snc, I-64100, Teramo, Italy \label{INAFOAA}
    }

   \date{}
   \titlerunning{Galaxy Evolution}
   \authorrunning{Golden-Marx et al. 2026}

  \abstract{Deep survey fields observed with JWST, particularly those hosting rich strong lensing galaxy clusters, offer unprecedented angular resolution and depth for studying galaxies in the early universe.  At cosmic noon, the environments of these galaxies (high-$z$ groups, clusters, and protoclusters) become vitally important as these structures typically host larger populations of star-forming galaxies.  By probing rich spectroscopic samples, it is possible to not only identify these structures spectroscopically, but also characterize the galaxies morphologically.  At 1 $< z <$ 4, these galaxies show irregular distributions, highlighting strongly star-forming clumps.}
  {Using predominantly recent JWST spectroscopic and photometric surveys, we search for structures at cosmic noon in the field surrounding Abell2744 and characterize the clumpy morphology of the galaxies in these structures.}
  {Accounting for lens corrected positions and magnitudes, we identify 11 structures at 1.0 $< z <$ 3.1.  Since we find primarily star-forming galaxies at all redshifts, we analyze the morphology of these galaxies to identify rest-frame U-band selected star-forming clumps to probe how these galaxies build their stellar mass. }
  {Among our 11 systems, half of our structures at $z <$ 1.5 are lower mass galaxy groups.  However, our five structures at $z >$ 2.4 may be better classified as protocluster cores.  In each structure, we identify primarily star-forming galaxies, which show strong agreement with the star-forming main sequence.  As part of our clump analysis, we find 61.8$\%$ of the galaxies in our sample have at least one off-axis clump. However, we find no strong trends with either redshift or the magnification of the galaxy.}  
  {By analyzing the population of clumps within each galaxy, we find statistically moderate trends linking the fraction of the luminosity encased in clumps to the total galaxy star formation rate, implying clumps are a dominant driver of star formation in protocluster galaxies at cosmic noon.  }


   \maketitle

\section{Introduction}\label{sect:intro}

Galaxy clusters are cosmic mega-cities that offer a unique environment to probe the evolution of galaxy properties as a function of redshift and environment.  Although low-$z$ clusters are characterized by large populations of quiescent, early-type galaxies \citep[e.g.,][]{Dressler1980,Miller2005,Rykoff2014,Rykoff2016, Vulcani+2023}, as we step back in cosmic time, cluster environments harbor increasing numbers of star-forming and merging galaxies \citep[e.g.,][]{Brodwin2013,Hennig2017}. However, while scaling relations characteristic of mature low-redshift clusters, such as the color-density relation, SFR-density relation, and morphology-density relation extend beyond $z$ $\sim$ 1 \citep[e.g.,][]{Postman2005,Nantais2016,Tomczak2019,Lemaux2019,Noordeh2021,Afanasiev2023,Mei2023}, there is growing evidence for younger galaxy populations with less well-defined properties \citep[e.g.,][]{Sazonova2020,Rinaldi2026} at these epochs.  Pushing further back to cosmic noon (2 $<$ $z$ $<$ 4), dense cluster-like environments are no longer fully virialized clusters, but rather protoclusters, the extended pre-cursor to present day galaxy clusters \citep[e.g.,][]{Muldrew2015,Overzier2016,Chiang2017,Alberts2022}, that simulations show will most likely gravitationally collapse into bound systems by $z$ = 0.  Observationally, protoclusters are identified as large, co-eval populations of galaxies spanning megaparsec scales and containing multiple, typically non-virialized, substructures \citep[e.g.,][]{Hatch2017,Cucciati2018,Shi2021,Forrest2023,Shah2024}.

The majority of identified protoclusters are dominated by actively star-forming galaxy populations \citep[e.g.,][]{Cucciati2018,Zheng2021,Shi2021,Edward2024,Forrest2024,Staab2024,Sikorski2026}. However, despite the high star-forming fractions typical of both protoclusters and field environments at these redshifts \citep[e.g.,][]{Muzzin2013A,Marchesini2014,Vulcani2016}, recent observations, particularly with JWST, have uncovered significant populations of quiescent galaxies in a growing number of early structures \citep[e.g.,][]{Willis2020,McConachie2022,Ito2023,McConachie2025,McConachie2025b,Toni2025b}.  However, comparing protocluster populations across different studies is often hindered by selection biases inherent to specific tracer populations.  Specifically, numerous protoclusters have been identified using single tracer types, such as Lyman-$\alpha$ emitters \citep[e.g.,][]{Cai2017,Umehata2019}; H-$\alpha$ emitters \citep[e.g.,][]{Zheng2021,Shi2021,Perez-Martin2023,Naufal2023,Golden-Marx2026}, or dusty star-forming galaxies \citep[e.g.,][]{Casey2015,Hill2020,Hill2024}.  As demonstrated in simulations \citep[e.g.,][]{Baxter2025} and in recent multi-wavelength follow-ups of these systems, broader galaxy selections and deep spectroscopy consistently reveal additional highly star-forming galaxies \citep{Zhang2022} as well as massive quiescent galaxies \citep[e.g.,][]{Shi2024,Naufal2024}.  These selection biases highlight the need for uniform population censuses, particularly through large, spectroscopically confirmed protoclusters that enable robust statistical comparisons across deep survey fields \citep[e.g.,][]{Lemaux2022,Forrest2024,Staab2024,Shah2024,Forrest2025,Hung2025,Gozaliasl2025,Toni2025a,Toni2025b,Sikorski2026}.  

A central question to protocluster science is how member galaxies evolve into the dominant, massive quiescent populations found in low-$z$ clusters \citep[e.g.,][]{DeLucia2007,vanDokkum2010,Oser2010,JGolden-Marx2022}.  As shown by \citet{Chiang2017}, at 2 $< z <$ 10, protoclusters account for 20 - 50$\%$ of the cosmic star formation rate density budget, with the most massive individual halos contributing $\sim$ 3$\%$ at cosmic noon.  These extreme levels of star formation activity are mirrored by recent findings of elevated stellar mass functions and enhanced star formation rates in early structures compared to the field \citep[e.g.,][]{Forrest2024,Staab2024,Sikorski2026}.  

Beyond their stellar mass growth, protocluster galaxies also look morphologically distinct from their low-$z$ counterparts, featuring an abundance of irregular, clumpy galaxies \citep[e.g.,][]{Peter2007,Naufal2023,Golden-Marx2026}.  While clump detection algorithms vary across studies, these star-forming clumps are typically UV-bright \citep[e.g.,][]{Kalita2025,Sok2026}, span pc-kpc scales \citep[e.g.,][]{Kalita2025}, and reach their peak frequency around cosmic noon \citep[e.g.,][]{Shibuya2016,Sok2026,delaVega2026}.  At $z$ $<$ 6, clump formation is largely attributed to in-situ processes: gas accretion fuels violent disk instabilities, driving gravitational collapse within turbulent disks \citep[e.g.,][]{Forster-Schreiber2009,Wisnioski2011,Wuyts2012,Forster-Schreiber2018,delaVega2026}.  This violent disk instability scenario is supported observationally by elevated gas mass fractions and turbulent, gaseous disks \citep[e.g.,][]{Daddi2010,Noble2017,Noble2019}.  Alternatively, clumps can form ex-situ via minor mergers and tidal interactions that compress gas into dense knots \citep[e.g.,][]{Guo2016,Ribeiro2017}. Evidence for ex-situ origins includes differences in clump metallicity compared to the host galaxy, implying some clumps are the remnants of previous galactic interactions \citep[e.g.,][]{Estrada-Carpenter2025,Sok2026a}.  

Despite multiple large clumpy galaxy studies, little has been done linking the environment to these properties.  At low redshift, galaxy environment drives galaxy morphology \citep[e.g.,][]{Dressler1980,Postman2005,Vulcani2011a,Vulcani2011b}. In protoclusters at cosmic noon, galaxy mergers are particularly important, as typical velocity dispersions ($\sim$ 300 - 500\,km\,s$^{-1}$) combined with high local densities significantly enhance interaction rates  \citep[e.g.,][]{Cucciati2014,Watson2019,Toshikawa2020,Liu2023,Liu2024,Puskas2025,Giddings2025}. Thus, it begs the question whether clumpy galaxies in protoclusters fundamentally differ from those in field environments.  To understand the role of clumpy star formation within cluster/protocluster galaxies, we undertake a study of spectroscopically-confirmed galaxies in overdense structures behind Abell2744.  

Regardless of their formation channel, robustly identifying and characterizing individual galaxy clumps requires high angular resolution.  As such, JWST has revolutionized this field \citep[e.g.,][]{Claeyssens2023,Kalita2024,Kalita2025,Kalita2025b,Kalita2025c,Claeyssens2025,Sok2026,delaVega2026,Zhu2026,Puskas2026}, allowing resolution scales previously only accessible via gravitational lensing.  While JWST resolves clumpy galaxies on $\sim$kpc scales, strong gravitational lensing and simulations show that individual clumps contain further substructure down to $\sim$pc scales \citep[e.g.,][]{Vanzella2022,Kalita2026-s1,Kalita2026-s2}, which will become accessible only with MICADO on the Extremely Large Telescope (ELT)\citep[e.g.,][]{Gilmozzi2007,Gilmozzi2008,Davies2016,Simioni2026}.  Here, we take advantage of the high angular resolution of the available JWST imaging combined with the strong gravitational lensing of Abell 2744 to characterize a phenomenological sampling of galaxies in dense structures at 1.0 $< z <$ 3.1, setting the first corner stone for exploring the clumpy properties of proto-structure galaxies with the maximum possible angular resolution currently available.

The paper is organized as follows.  We introduce our sample of spectroscopically confirmed galaxies, as well as the underlying galaxy photometry and clump methodology in Section~\ref{sect:data}.  We present our results regarding the galaxy populations in our structures and the clumpiness of these galaxies in Section~\ref{sect:results-M}.  We contextualize our results across the broader literature of high-$z$ structures and clumpy galaxies in Section~\ref{sect:Discussion}.  Lastly, we present our conclusions and plans for future work in Section~\ref{sect:Conclusions}.  Throughout this paper, we adopt a flat $\Lambda$CDM cosmology, using H$_{0}$ = 70\,km\,s$^{-1}$\,Mpc$^{-1}$, $\Omega_{m}$ = 0.3, and $\Omega_{\Lambda}$=0.7.  Unless otherwise noted, all distances are given as proper distances, and all magnitudes are given as AB magnitudes.  

\section{Data $\&$ Analysis}\label{sect:data}
\subsection{Spectroscopic Sample}\label{sect:spec-z}
Abell 2744  is one of the most well-studied strong gravitational-lensing cluster fields \citep[e.g.,][]{Zitrin2014,Mahler2018,Steinhardt2020,Bergamini2023a,Bergamini2023b,Furtak2023}.  As a result, both the cluster and its surrounding field have been extensively surveyed, with recent JWST observations providing detailed characterization of thousands of galaxies \citep[e.g.,][]{Treu2022,Treu2023,Wang2023,Merlin2024,Naidu2024,Bezanson2024,Suess2024,Wang2024,Weaver2024,Price2025,Watson2025}, including multiple group/cluster/protocluster candidates \citep[e.g.,][]{Morishita2023,Wu2023,Watson2025,Pan2025,Li2026}.

To identify and characterize structures at cosmic noon in a uniform manner, we first need to establish a robust and comprehensive sample of spectroscopically confirmed galaxies.  To this end, we compiled redshift measurements from the extensive spectroscopic datasets available in the field, including GLASS-JWST using the Near-Infrared Spectrograph (NIRSpec) \citep{Treu2022,Mascia2024}, UNCOVER and Medium Bands: MegaScience using NIRSpec \citep{Bezanson2024,Suess2024,Weaver2024,Price2025}, Near-Infrared Camera Wide-field Slitless Spectroscopy (NIRCam WFSS) \citep{Naidu2024}, GLASS-JWST Near-Infrared Imager and Slitless Spectroscopy (NIRISS) \citep{Watson2025}, as well as existing archival ground-based observations and MUSE data \citep[][]{Couch1987,Couch1998,Boschin2006,Braglia2009,Owers2011,Schmidt2014,Treu2015,Richard2021,Torralba2026}. This compilation provides a catalog of all available spectroscopically-confirmed galaxies in the field, enabling a homogeneous identification of potential structures. 

Because each catalog used different detection mechanisms for their spectral classification, as well as different segmentation maps for matching to photometric galaxies, our scheme for creating a single spectroscopic catalog is described below.  
\begin{itemize}
    \item For each catalog, we only kept galaxies with secure spectroscopic measurements.  For the GLASS-JWST NIRSpec sample, we followed the recommendations of \citet{Mascia2024} and only used galaxies with a spectral quality flag of 3 or 4, which signify a very reliable redshift, with an estimated accuracy of 99$\%$ and at least one clear spectral feature.  For the UNCOVER/MegaScience sample, we followed \citet{Price2025,Bezanson2024} and required our secure spectra to have a spectral quality flag of 2 or 3, which corresponds to at least one secure spectral feature. For the NIRCam WFSS catalog sample, we followed \cite{Naidu2024} and classified secure spectra as those having at least two emission lines with a S/N $>$ 5.  For the GLASS-JWST NIRISS, we followed \citet{Watson2025} and only used spectra with a spectral quality flag of 4, corresponding to secure spectra with multiple features or lines.  As these secure spectra include predominately star-forming galaxies due to the required emission lines, we address potential biases in our cluster detection in Section~\ref{sect:quiescent} and Appendix~\ref{Appendix}.  
    \item We used the GLASS-JWST NIRSpec catalog \citep{Mascia2024} as our primary source and 
    matched it with the JWST UNCOVER NIRSpec catalog \citep{Price2025}.  We adopted a matching radius of 0$\farcs$9, chosen to maximize the number of matches while minimizing multiple matches. We verified the robustness of this choice by varying the radius between 0$\farcs$6 and 3$\farcs$0;   after removing sources with multiple counterparts, the resulting spectroscopic sample remained unchanged.  In the event of multiple sources from the UNCOVER catalog matching with the same source from the GLASS-JWST NIRSpec, we kept the closest match and flagged the galaxy as some of these may represent multiple galaxy components that have spectroscopic redshifts.  Galaxies in the UNCOVER catalog without a match in the GLASS-JWST NIRSpec catalog were added to the final catalog.    
    \item We repeated this process with both the NIRCam WFSS catalog \citep{Naidu2024} and the GLASS-JWST NIRISS catalog \citep{Watson2025}, using identical matching criteria, and similarly removed any double matches and added unmatched galaxies to the final sample.  
    \item Lastly, we matched the combined JWST spectroscopic catalogs with separate archival spectroscopic catalogs built using ground-based spectroscopic catalogs and MUSE data \citep[][]{Couch1987,Couch1998,Boschin2006,Braglia2009,Owers2011,Schmidt2014,Treu2015,Richard2021,Torralba2026}.  Similar to the other catalogs, we matched galaxies within 0$\farcs$9.
    \item Based on the combined catalog, we selected a single redshift for each galaxy.  For galaxies with multiple redshift matches, we prioritized the redshift based on the order of the catalog matching (e.g., GLASS-JWST NIRSpec, UNCOVER, NIRCam WFSS, GLASS-JWST NIRISS, ground-based/MUSE). 
    Although the overwhelming majority of galaxies have well-matched spectroscopic redshifts between catalogs, there were a small number of sources, primarily matches with older ground-based redshifts, that differ by $\Delta$z $>$ 1.0.  While some of these represent high-$z$ galaxies outside of our target window ($z$ $>$ 7.0), there are a handful of cases closer to cosmic noon where the ground-based redshifts do not agree with the JWST redshifts.  In these cases, we kept the JWST redshifts over the initial ground-based redshifts.  
\end{itemize}

To build our final science catalog, we matched each spectroscopic galaxy to a photometric galaxy from UNCOVER \citep[e.g.,][]{Wang2023,Wang2024,Suess2024,Weaver2024}.  Although we removed duplicates from our spec-$z$ catalog, this added step prevents the double counting of multi-component, clumpy galaxies, which peak at cosmic noon \citep[e.g.,][]{Shibuya2016,Sok2026,delaVega2026,Puskas2026}.  Although the number of galaxies with two distinct spectroscopic detections within the same photometric footprint is low, a number of our galaxies are characterized as clumpy, multi-component systems (see Section~\ref{sect:Clump-Dection}).  However, by matching to UNCOVER photometry, we also removed all spectroscopic galaxies outside of the UNCOVER footprint (primarily from our ground-based and MUSE catalogs).  

\begin{figure}
\begin{center}
\includegraphics[scale=0.55,trim={0.0in 0.0in 0.0in 0.5in},clip=true]{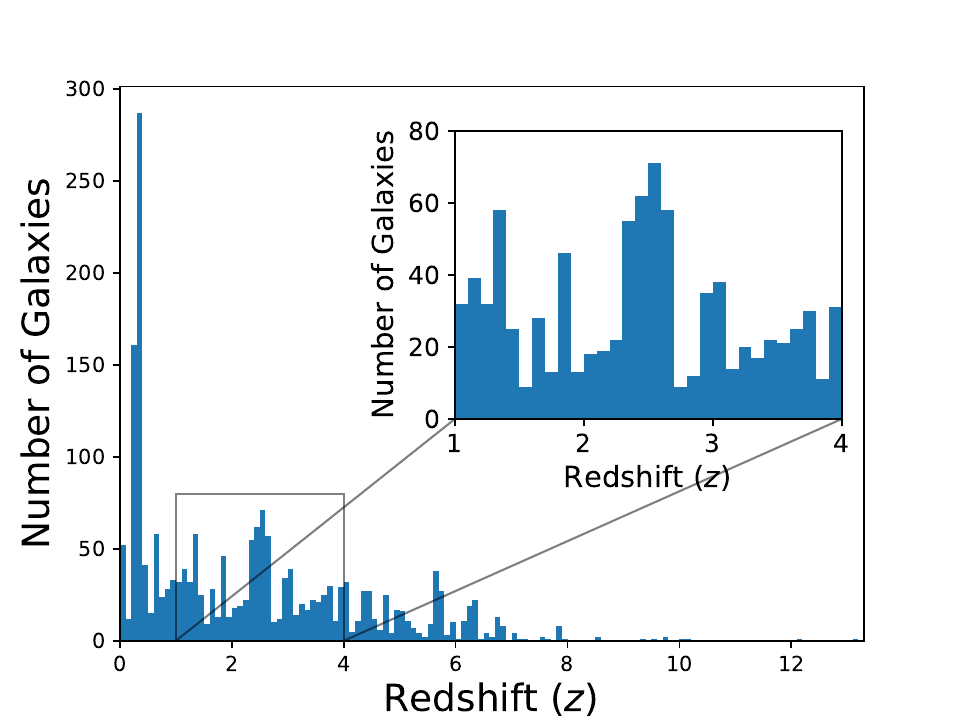}

\caption{Main: Redshift histogram of our complete spectroscopic catalog for the Abell 2744 field.  The redshift of Abell 2744 is 0.3072 \citep{Bergamini2023a}.  Inset: Zoomed in histogram of the 885 galaxies 1.0 $\leq z \leq$ 4.0, the redshift range of interest in this study.}

\label{Fig:Specz-Hist1}
\end{center}
\end{figure}

The final spectroscopic catalog includes 2007 unique galaxies at 0.0 $< z < $ 13.2 (see Figure~\ref{Fig:Specz-Hist1}).  However, because the field surrounding Abell 2744 is strongly lensed, this sample includes strong lensing arcs and double images.  While some of these sources have unique spectra, they act as interlopers for any potential structure identification.  As such, we removed multiple images only from structures identified after undertaking an analysis of which spectroscopically-confirmed galaxies were statistically most-likely to host multiple images based on our lens modeling (see Section~\ref{sect:Lensing} and \ref{sect:Clusters}).

\subsection{Lensing Corrections}\label{sect:Lensing}
Although the inset histogram in Fig.~\ref{Fig:Specz-Hist1} shows peaks in the distribution of galaxies at cosmic noon, clusters require spatial and redshift overdensities.  As such, we must correct the positions and the magnitudes of each spec-$z$ galaxy in our sample to accurately identify any overdensities. For this analysis, we used a new strong gravitational lensing model constructed with \texttt{Gravity.jl} \citep{Lombardi2024} based on the model presented in \citet{Bergamini2023a,Bergamini2023b}.  This model was cross-calibrated to optimize consistency in cluster mass parameters and derived magnifications ($\mu$) and is fully presented in Di Rosa et al. (in prep).  To correct the positions and magnifications of each galaxy in our sample, we randomly selected 1000 iterations from the 16000 Markov Chain Monte Carlo (MCMC) iterations generated by the lensing model to compute the source-plane positions and lensing magnifications of our targets.  As a result, the new positions of each galaxy are the median RA and Dec reported for these 1000 MCMC iterations.  Similarly, we used the median magnification output for each iteration.  To correct the reported magnitudes from the UNCOVER catalogs for our new measurements of magnification, we added a magnitude correction factor (2.5\,log$_{10}$($\mu$)) when converting aperture fluxes from UNCOVER into magnitudes \citep[e.g.,][]{Wang2023,Wang2024,Suess2024,Weaver2024}.  For the remainder of this paper, all magnitudes and galaxy positions refer to the lens-corrected galaxies.  We also used the best-fit lens model obtained from the MCMC chains to determine which sources in our target catalog are expected to be multiply imaged (see Section~\ref{sect:DoubleImages}).      

\subsection{Identifying overdense structure candidates}\label{sect:Clusters}
To identify structures at cosmic noon, we limited our analysis to the 885 galaxies at 1.0 $\leq$ $z$ $\leq$ 4.0.  Despite the robust spectroscopic sample and lensing-corrected galaxy positions and magnitudes, our spectroscopic selection is intrinsically non-uniform.  The input catalogs stem from a combination of targeted observations (e.g., GLASS-JWST), and slitless-grism redshifts (e.g., NIRCam WFSS and NIRISS).  To ensure spatial and flux homogeneity and identify structures uniformly, we limited our analysis to spectroscopically confirmed galaxies brighter than a locally determined magnitude limit, based on the photometric counterparts of the spectroscopic sources (see Table~\ref{tb:clusters}). 

Given the non-uniform sampling of our spectroscopic catalog, we determined a separate magnitude limit around each prominent peak in the redshift distribution (see Fig.~\ref{Fig:Specz-Hist1}), accounting for the local spectroscopic completeness.  
For the majority of our sample, we derived a lensing-corrected limit of $m_{\mathrm{F150W}} = 25\,\mathrm{mag}$, which we adopted for our initial structure detection limit (see Appendix~\ref{Appendix} for a discussion of the impact of the magnitude limits and spectroscopic  completeness).  For the highest redshift structures $z >$ 2.9, the limit shifts to m$_{F150W}$ = 27\,mag, while for the structures at $z$ $\sim$ 1.10 and 1.87, we adopted m$_{F150W}$ = 26\,mag (see Table~\ref{tb:clusters} for individual structure limits).  All threshold cuts are significantly brighter than the 5$\sigma$ depth of $28.35\,\mathrm{mag}$ reported by \citet{Wang2023}, ensuring high photometric completeness across our spectroscopic matching.

Using this magnitude-limited sample, we searched for candidate structures using the full 3D spatial galaxy distribution.  Specifically, we measured the density by counting the number of galaxies within a cylindrical volume centered on each spectroscopic galaxy. Following \citet{Calvi2021,Helton2024a,Helton2024b}, we adopted a cylinder radius of 1\,Mpc (evaluated at the central galaxy's redshift) and line-of-sight velocity window of $\Delta v = \pm 1000\,\mathrm{km\,s^{-1}}$.  These parameters encompass the typical spatial extent (0.5 -- 1\,Mpc; \citealp[e.g.,][]{Wylezalek2013,Cooke2015,Paterno-Mahler2017,Golden-Marx2019,Gonzalez2019,Golden-Marx2021,Thongkham2024}) and the velocity dispersions ($\sim$ 1000\,km\,s$^{-1}$ \citealp[e.g.,][]{Noirot2016,Noirot2018}) of high-$z$ clusters and protoclusters ($\sim$ 500\,km\,s$^{-1}$ \citealp[e.g.,][]{Cucciati2014,Toshikawa2020,Liu2024,CWatson2025}).  However, at $z$ $>$ 2, a 1\,Mpc region corresponds to the typical size of a protocluster core as protoclusters often span significantly larger spatial scales (5 - 6.5\,Mpc) and contain multiple density peaks \citep[e.g.,][]{Cucciati2018,Shi2021,Forrest2023,Staab2024,Shah2024}.  While these selection criteria successfully isolate the highest-density peaks in the structure distribution, they do not fully map the extended outskirts, where galaxy overdensities blend into the field.

To better determine which structures are cluster candidates, we account for redshift-dependent variations in the median number of galaxies by calculating the median density of galaxies in sliding redshift slices ($\Delta$z $\sim$ $\pm$ 0.1) around each potential density peak.  We divided the sample into potential structures and co-eval field galaxies by isolating galaxies above the median galaxy density as structure candidates (see Figure~\ref{Fig:Cluster-Density-Plot}).  This analysis identified 14 structures at 1.0 $< z <$ 3.1 with density peaks above the local background measurement, which will be further studied in Section~\ref{sect:results-clusters}. 

\subsection{Spectral Energy Distribution Modeling}\label{Sect:SED}
To evaluate the stellar properties of our sample homogeneously, we performed custom spectral energy distribution (SED) modeling using CIGALE \citep{Boquien2019,Yang2020}, adopting our spectroscopic redshifts across $1.0 \leq z \leq 4.0$. While public UNCOVER catalogs provide Prospector-$\beta$ SED fits \citet{Wang2024}, those relied on a smaller spec-$z$ baseline and yielded non-negligible photo-$z$ discrepancies for a subset of our targets.  Using the same 0$\farcs$32 aperture photometry from UNCOVER \citep[e.g.,][]{Wang2023,Weaver2024,Wang2024}, we follow parameter configurations similar to \citet{Shen2021} and \citet{Shen2025} (see Table~\ref{tb:CIGALE}).  We fit all available UNCOVER photometric bands (HST: F435W, F606W, F814W, F105W, F125W, F140W, F160W; JWST: F070W, F090W, F115W, F140M, F150W, F162M, F182M, F200W, F210M, F250M, F277W, F300M, F335M, F356W, F360M, F410M, F430M, F444W, F460W, F480M) and adopted a \citet{Chabrier2003} Initial Mass Function (IMF) and a \citet{Bruzual2003} Stellar Synthesis Population Model.  For galaxies where our spec-$z$ aligns with the UNCOVER photo-$z$, our derived stellar masses and star formation rates (SFRs) show agreement with published values \citep[e.g.,][]{Wang2023,Wang2024}.

\begin{table*}[h!]
    \caption{
    Model components used for the SED fitting with CIGALE.}
    \label{tb:CIGALE}
    \centering
    \begin{tabular}{lp{5.5cm}p{7cm}}
        \hline
        \multicolumn{3}{c}{\textbf{Star Formation History}}\\
        \hline
        \multicolumn{3}{c}{parametric:\textit{ sfhdelayed}}\\
        \hline
 $\tau_{\mathrm{main}}$ [Myr] & 100, 300, 500, 1000, 3000 & e-folding time of main stellar population in Myr\\ 
        age [Myr] & 300, 500, 750, 1000, 2000, 3000 & Age of main stellar population in Myr\\
        $\tau_{\mathrm{burst}}$ [Myr] & 50 & e-folding time of late starburst population model\\
        age$_{\mathrm{burst}}$ [Myr] & 10, 20, 50 & Age of late starburst in Myr\\
        f$_{\mathrm{burst}}$ & 0.0, 0.01, 0.05, 0.1 & Mass fraction of the late starburst population\\
        \hline
        \multicolumn{3}{c}{\textbf{Simple Stellar Population: BC03}}\\
        \hline
        IMF & Chabrier & Initial mass function \\
        Z & 0.004, 0.008, 0.02 & Metallicity \\   
        \hline
        \multicolumn{3}{c}{\textbf{Attenuation: dustatt\_modified\_starburst}}\\
        \hline
         E(B - V) & 0.01, 0.05, 0.1, 0.15, 0.2, 0.25, 0.3, 0.4, 0.5, 0.6, 0.8, 1.2 & Color excess of nebular lines for both stellar populations  \\        
    \end{tabular}
    \tablefoot{
    Any parameters not explicitly listed were left as the default values.}\\
\end{table*}

We adopt the $100\,\mathrm{Myr}$ time-averaged star formation rate ($\mathrm{SFR}_{100}$) and total stellar mass ($M_*$), together with their associated parameter uncertainties. In addition, we derive rest-frame absolute $U$-, $V$-, and $J$-band fluxes with CIGALE, enabling a uniform characterization of galaxy colors and quiescent fractions as a function of redshift (Section~\ref{sect:results-clusters}). All derived physical quantities - including stellar masses, SFRs, and absolute magnitudes - are explicitly de-magnified using our lensing model, consistent with the treatment of our photometric depth limits.

\subsection{Characterizing Clumpy Galaxies}\label{sect:Clump-Dection}
To detect star-forming clumps in galaxies, we implemented a smoothed subtraction technique similar to \citet{Guo2016,Kalita2024,Kalita2025} and building upon the methodology presented in Giddings et al. (in prep).  Clump identification was performed in the filter tracing the rest-frame U-band for each target (see Table~\ref{tb:clusters}).  Although we traced rest-frame U-band for each structure, the width and the position of the rest-frame U-band do differ greatly between wavebands and structures (see Table~\ref{tb:clusters}).  Specifically, the majority of our structures are detected in wide bands (F070W, F090W, F115W, F150W), while the measurements of clumpy galaxies in the structure at $z$ = 2.66 was done in the F140M band.  Furthermore, we used the F115W band to characterize galaxies at 1.87 $\leq$ $z$ $\leq$ 2.58, meaning that the response function of the filter and the wavelengths the filter traces does differ throughout the analysis.  As such, some clump variations, particularly regarding clump magnitude limits and trends with SFR discussed in Section~\ref{sect:Clump-results}, could be due to these differences.  

\begin{figure*}
\begin{center}
\includegraphics[scale=0.21,trim={9.5in 2in 48.in 1.4in},clip=true]{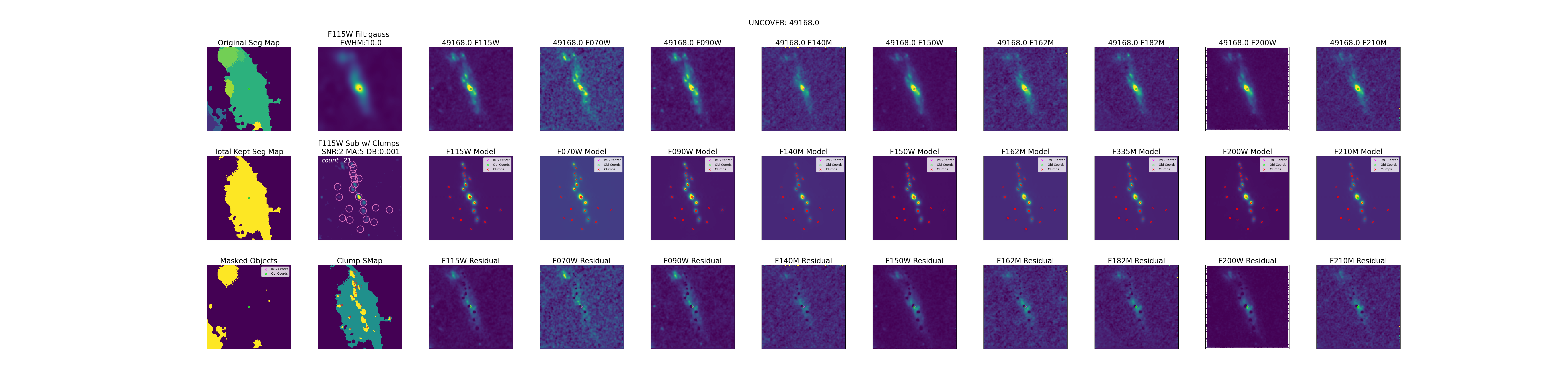}
\vrule width 0.5pt
\includegraphics[scale=0.21,trim={9in 2in 48in 1in},clip=true]{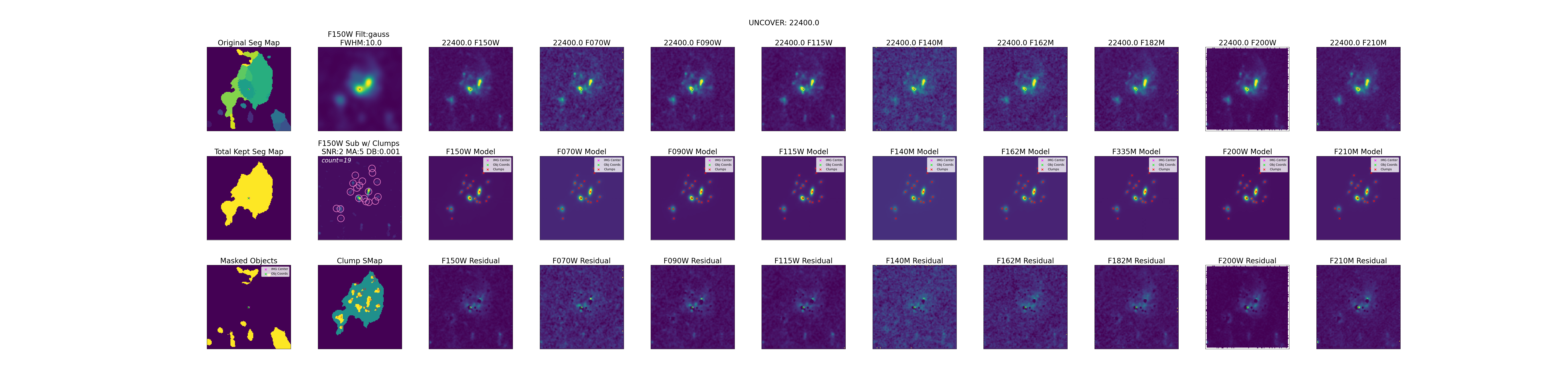}
\caption{5$\arcsec$ $\times$ 5$\arcsec$ cutouts exemplifying our clump detection method.  The detection was done in the rest-frame U-band (F115W in the left panel set at $z$ = 2.502, F150W in the right panel set at $z$ = 3.059).  For each example, the left column shows the set of segmentation maps for each target galaxy.  The top left shows each segmentation map within our cutouts, while the middle left shows the total segmentation map area used for our clumps detection, which includes the initial target galaxy and any adjacent galaxies at a similar redshift.  The bottom left panel shows all masked sources in the field.  The top right panel shows the initial science image, while the top middle panel shows the smoothed image.  The middle panel shows the subtracted images, with all clumps identified in purple circles.  The count shown in the image accounts for our clump detection before the clump magnitude limit is applied. The middle right panel shows our model of the clumps, with each clump marked by a red x. The bottom middle panel shows the segmentation map of each clump and the bottom right panel shows the residual image after clump subtraction.}

\label{Fig:Clumps-Example}
\end{center}
\end{figure*}

For every galaxy, we constructed a 5$\arcsec$ $\times$ 5$\arcsec$ cutout centered on the UNCOVER photometric coordinates (see the top right panel in each of the two examples in Fig.~\ref{Fig:Clumps-Example}) and restricted our analysis to the UNCOVER segmentation map (see the left-hand panels in the two examples in Fig.~\ref{Fig:Clumps-Example}).  Prior to detection, the BCG-subtracted UNCOVER images were smoothed to the same angular resolution and re-scaled to match the pixel scale of the F444W PSF-matched catalog, upon which the UNCOVER source detection and segmentation maps are defined \citep{Suess2024}.  Finally, to ensure accurate noise estimation across the cutouts, we evaluated the signal-to-noise ratio (S/N) in blank-sky regions and rescaled the error maps such that the standard deviation of the regions without UNCOVER detected sources had a $\mathrm{S/N}$ distribution equal to unity.

Next, we identified clumps by applying a Gaussian filter (with a FWHM of 10 pixels [$\sim$ 0$\farcs$4]) to generate a smoothed host-galaxy model (see the top middle panel of each example in Fig.~\ref{Fig:Clumps-Example}). At 1 $<$ $z$ $<$ 3.1, this kernel corresponds to a physical scale of $\sim$ 3\,kpc, which is larger than the typical $\sim$1\,kpc radius of a non-lensed high-$z$ clump \citep[e.g.,][]{Kalita2025,Kalita2025b,Sok2026}.  As such, this smoothing blends the small-scale structure into the background galaxy.  We then subtracted the smoothed model from the original science image to create a difference map used for our initial clump extraction (see the middle panel of each example in Fig.~\ref{Fig:Clumps-Example}). Before creating our clump catalog, we applied a mask to remove any potential clumps with a S/N $<$ 2.  We then used SEP \citep[][]{Bertin1996,Barbary2016}, a python implementation of SOURCE EXTRACTOR, to detect all remaining clumps.  The detected parameters  (e.g., the pixel coordinates of the galaxy, semi-major axis (a), S\'ersic index (n), ellipticity, and orientation angle) are then passed as fixed strict priors with small, well defined boundaries, to our PYSERSIC fitting \citep{Pasha2023} of the full galaxy image to model the light profile and measure the total clump flux.  Although our output catalogs include r$_{\mathrm{eff}}$ and ellipticity, these values are set by our SEP detection, with r$_{\mathrm{eff}}$ being a PSF conversion of the SEP measured semi-major and minor axes.  To account for spatial resolution limits, we separated candidate features into resolved clumps, (modeled with a Sérsic profile) and unresolved sources ($<$ 1.5\,r$_{\mathrm{psf}}$; modeled as point sources).  

Because the initial difference map slightly over-subtracts clump flux and underestimates the local background due to diffuse host-galaxy emission, we followed \citet{Kalita2025} and measured the diffuse galactic light in local elliptical annuli surrounding each clump. We scaled the host-galaxy contribution over the estimated clump footprint ($2\,r_{\mathrm{eff}}$) and adjusted the net clump fluxes accordingly.

Because clumpy galaxies can be associated with enhanced galaxy-galaxy merger activity \citep{Ribeiro2017}, we explicitly accounted for potentially-interacting neighbors within the UNCOVER footprint by comparing their photometric redshift probability distribution functions (PDFs) to the spectroscopic redshifts of each target galaxy.  We required these neighbors to ``touch", such that they are within 2$\arcsec$ of the host galaxy and have segmentation maps adjacent to the target galaxy (see the middle left panel of Fig.~\ref{Fig:Clumps-Example}). For each photometric neighbor, we modeled the redshift PDF as an asymmetric Gaussian defined by the $z_{16}$ and $z_{84}$ quantiles from UNCOVER \citep{Wang2024}.  We integrateded this PDF over a window of ($\pm$ 0.05\,[1 + $z_{\mathrm{spec}}$]) and retained any neighbor with a $>1\%$  probability of residing at the target redshift .  We intentionally adopted a permissive threshold to avoid missing interacting systems, given that larger photo-$z$ uncertainties at higher redshift complicate stricter selections \citep[e.g.,][]{Golden-Marx2026}.  In practice, most nearby photometric neighbors are low-mass systems with negligible impact on the host galaxy.  In cases where two spectroscopically confirmed cluster members are spatial close-pairs (see Section~\ref{sect:Clump-results}), we treated the pair as a single merged system for clump identification.

To ensure uniform clump population statistics across the sample, we derived clump detection magnitude limits for each band used.  Since clump detectability depends on both the local noise floor and the diffuse background light of the surrounding galaxy, these limits were calculated on a target-by-target basis, where we examined the distribution of clump magnitudes of a given cluster. Each limit is de-magnified using the galaxy's gravitational lensing magnification.  Although magnification varies spatially across extended targets, modeling pixel-scale lensing gradients is beyond the scope of this paper; thus, we adopted a single magnification factor per galaxy calculated at its central sky coordinates.  As shown in Table~\ref{tb:clusters}, after magnification corrections,  our clump detection limits range from 28.0\,mag in F090W to 30.0\,mag in F150W.  While these limits align with the nominal UNCOVER survey depth \citep[e.g.,][]{Weaver2024,Suess2024}, the compact angular size of typical clumps ($\sim$ 0$\farcs$05) allows for reliable detection compared to extended sources evaluated in larger photometric apertures ($\sim$ 0$\farcs$32 in \citep{Weaver2024}; see Fig.~\ref{Fig:Clumps-Radius}).   

\subsubsection{Proto-bulge vs star-forming clumps}
As shown in Figure~\ref{Fig:Clumps-Example}, detected clumps can encompass physically distinct components. In particular, bright central clumps often correspond to galaxy bulges-in-formation (proto-bulges hereafter).  Following \citet{Sok2026}, we isolated these central features by defining a proto-bulge as any clump detected within $0\farcs2$ of the host galaxy's photometric centroid. For the few systems hosting multiple candidate proto-bulges within this search radius, both are classified as proto-bulges and removed them from the clump population.  Although minor contamination may persist, particularly in complex merging systems with combined segmentation footprints, the clump counts presented throughout the remainder of this work explicitly refer to the population excluding central proto-bulges.

\section{Results}\label{sect:results-M}

\subsection{Identifying \& Characterizing Cluster/Protocluster Candidates}\label{sect:results-clusters}

\begin{figure}
\begin{center}
\includegraphics[scale=0.45,trim={0.3in 0.0in 0.1in 0.6in},clip=true]{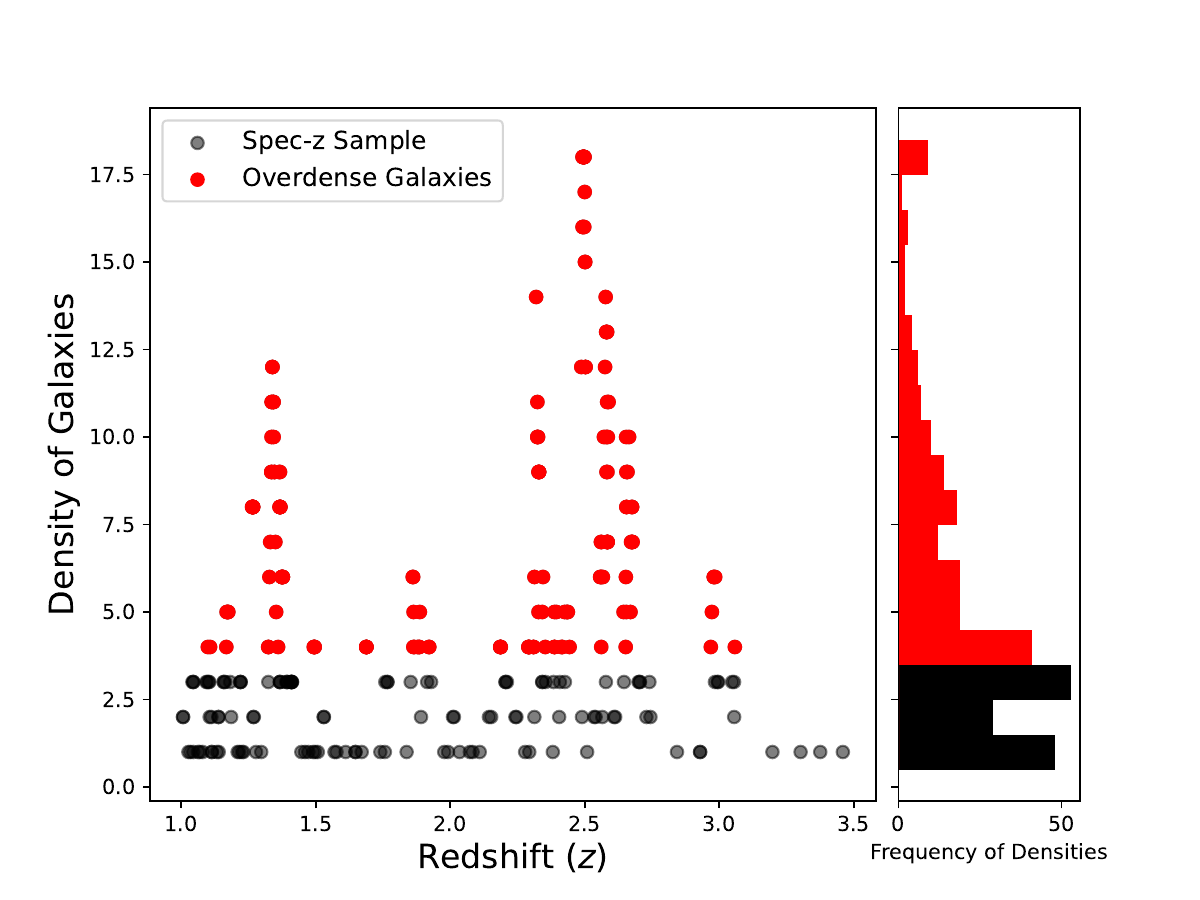}

\caption{Density of spec-$z$ galaxies.  Left: The density as a function of galaxy redshift. The density shown is the number of spec-$z$ galaxies within our cylindrical search region centered on each spec-$z$ galaxy. The red points are galaxies with densities above the median for the entire redshift sample, while the black points are the galaxies below the median density.  These peaks represent the 14 initial structure candidates in the Abell 2744 field. Right: Histogram of the frequency of densities. The same red/black divide as in the left panel highlights the elevated fraction of galaxies in low density environments (below the median value). }

\label{Fig:Cluster-Density-Plot}
\end{center}
\end{figure}

\begin{figure*}
\begin{center}
\includegraphics[scale=0.45,trim={0.0in 0.0in 0.0in 1.2in},clip=true]{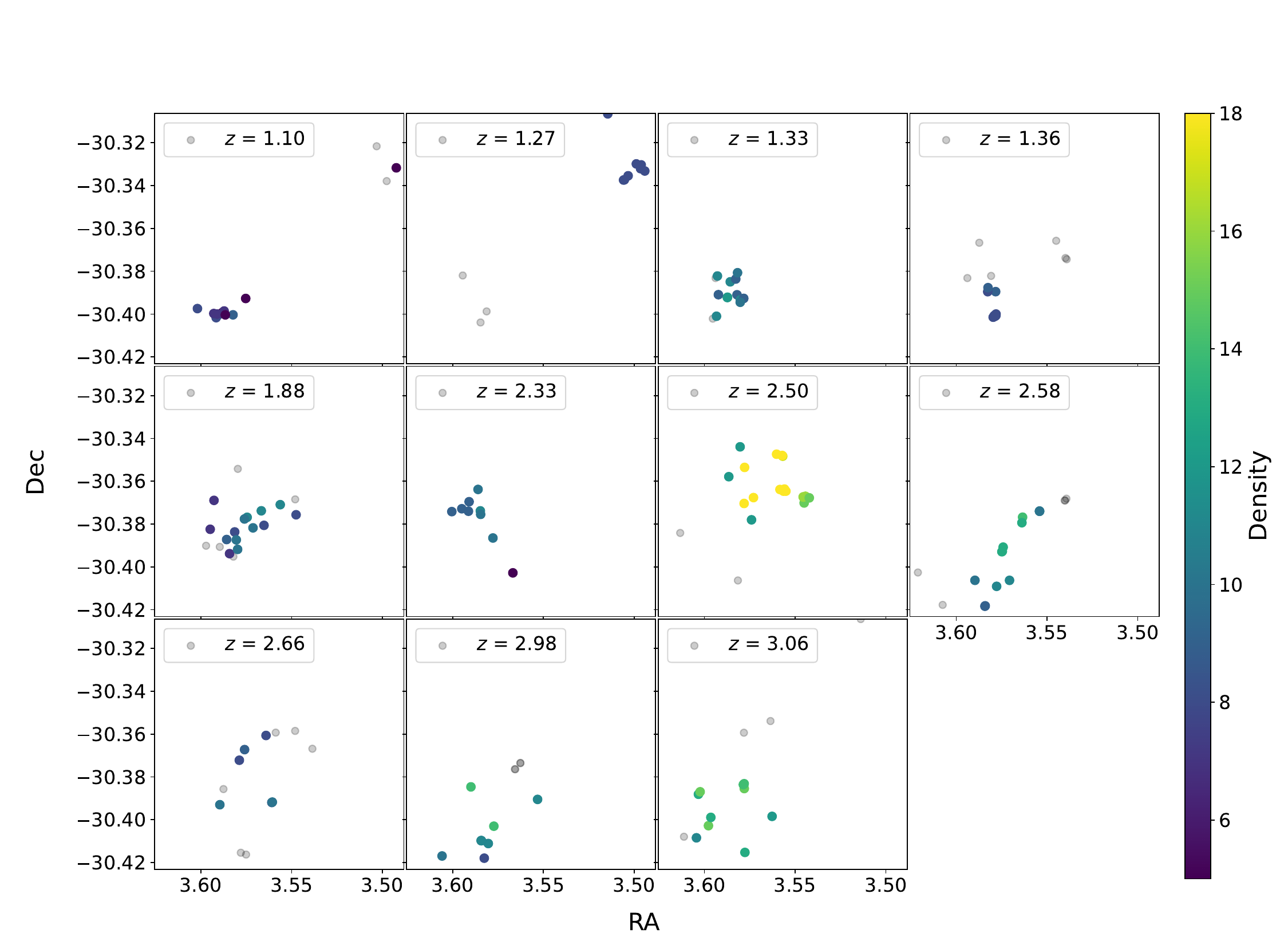}

\caption{Density maps of spec-$z$ galaxies shown over the entire Abell 2744 field.  We plot the de-lensed position of the spec-$z$ galaxies at each redshift epoch in Fig.~\ref{Fig:Cluster-Density-Plot}. The colored points show our cluster members as defined in Section~\ref{sect:Clump-Dection}, while the gray points are co-eval galaxies that we do not classify as being cluster members either because of their density.  The density color bar scale is based on the number of galaxies within the discussed cylindrical search region.  Although we use a uniform color and scale bar, the minimum threshold for being in a structure changes slightly depending on the redshift.}

\label{Fig:DensityMaps}
\end{center}
\end{figure*}

Although our initial 3D cylindrical structure detection yielded 14 overdense peaks (see Section~\ref{sect:Clusters}), we followed standard identification methods for overdense structures \citep[e.g.,][]{Wylezalek2013, Paterno-Mahler2017, Golden-Marx2019} and refined this sample by requiring an overdensity significance $>$3 above the median density value, our estimate of a local background (see the right panel in Figure~\ref{Fig:Cluster-Density-Plot}). Additionally, because all structures are found within a relatively small, gravitationally-lensed field-of-view (see Fig.~\ref{Fig:DensityMaps}),  we imposed a minimum rest-frame velocity separation of 3000\,km\,s$^{-1}$ between peak centroids to distinguish distinct structures from sub-components of potentially merging systems.  These criteria yield a final sample of 11 structures (see Table~\ref{tb:clusters} and Fig.~\ref{Fig:Cluster-Density-Plot}), including one potentially merging double-peaked structure at $z$ = 1.87.  Although we treat these 11 systems as distinct entities throughout this paper, the two structures at $z$ = 1.34 and 1.37 are separated by small spatial and velocity offsets just beyond our merger threshold and may also constitute an ongoing cluster-cluster merger.  Similarly, the three structures at 2.5 $\leq$ $z$ $\leq$ 2.66 and the two structures at $z =$ 2.98 and $z =$ 3.05 could each be parts of broader, contiguous protocluster complexes. Such extended architecture is reminiscent of massive structures like Hyperion (2.41 $< z <$ 2.68) \citep[e.g.,][]{Cucciati2018,Forrest2025,Giddings2025,Sikorski2026}, Elentari (3.20 $< z <$ 3.45) \citep{Forrest2023}, and Smruti (3.38 $< z <$ 3.54) \citep{Shah2024}. Determining whether these individual peaks are physically bound within overarching super-structures would require a full probabilistic treatment of photometric redshift populations \citep[e.g.,][]{Lemaux2018,Hung2025} convolved with the lensing models, which is beyond the scope of this work.  
 
\begin{table*}[h!]
\caption{Detected High-$z$ Structure Properties}                 
\label{tb:clusters}   
\centering                      
\begin{tabular}{l c c c c c c c c c}     

\hline\hline             

Cluster & \multicolumn{2}{c}{Central\tablefootmark{a}} & Redshift & Galaxy & Number & $\sigma_{\parallel}$\tablefootmark{c} & Clump & Rest-frame  & Clump\\ 
& \multicolumn{2}{c}{Coordinates} & & mag. limit & of spec-$z$ &(km\,s$^{-1}$) & detection & Wavelength & mag limit\\
& RA & Dec & & (mag, (m*)\tablefootmark{b}) & members &  & band & ($\mu$m) & (mag)\\

\hline
Cluster 1 & 3.586 & -30.397 & 1.10 & 26.0 (m*+4.7) & 11 & 832.5 $\pm$ 162.2 & F070W & 0.34 & 29.0\\
Cluster 2 & 3.498 & -30.332 & 1.27 & 25.0 (m*+3.4) & 8 & 176.9 $\pm$ 65.4 & F090W & 0.40 & 28.0\\
Cluster 3 & 3.587 & -30.391 & 1.34 & 25.0 (m*+3.3) & 12 & 581.6 $\pm$ 77.8 & F090W & 0.39  & 29.5\\
Cluster 4 & 3.580 & -30.400 & 1.37 & 25.0 (m*+3.3) & 8 & 194.6 $\pm$ 65.9 & F090W & 0.38 & 29.0\\
Cluster 5 & 3.580 & -30.375 & 1.87 & 26.0 (m*+3.6) & 14 & 1251.5 $\pm$ 253.1\tablefootmark{d} & F115W & 0.40 & 29.5\\
Cluster 6 & 3.578 & -30.387 & 2.33 & 25.0 (m*+2.2) & 9 & 248.1 $\pm$ 37.4 & F115W & 0.35 & 28.0 \\ 
Cluster 7 & 3.557 & -30.365 & 2.50 & 25.0 (m*+2.0) & 18 & 383.0 $\pm$ 61.2 & F115W & 0.33  & 29.0\\
Cluster 8 & 3.573 & -30.941 & 2.58 & 25.0 (m*+1.9) & 13 & 381.8 $\pm$ 101.2 & F115W & 0.32 & 29.5\\
Cluster 9 & 3.580 & -30.384 & 2.66 & 25.0 (m*+1.9) & 5 & 168.9 $\pm$ 109.3 & F140M & 0.38 & 28.5\\
Cluster 10 & 3.578 & -30.405 & 2.98 & 27.0 (m*+3.6) & 9 & 335.8 $\pm$ 115.5 & F150W & 0.38 & 28.0\\ 
Cluster 11 & 3.580 & -30.385 & 3.05 & 27.0 (m*+3.5) & 10 & 195.3 $\pm$ 52.1 & F150W & 0.37 & 30.0\\

\hline                                
\end{tabular}
\tablefoot{\tablefoottext{a}{Mass weighted, lens-corrected central coordinates of our structures}\\
\tablefoottext{b}{The magnitude limit relative to the apparent magnitude of an L* galaxy estimated using EzGal \citep{Mancone2012}.}\\
\tablefoottext{c} {These values are estimated using the Gapper method with jackknifed confidence intervals, following \citet{Beers1990}}\\
\tablefoottext{d} {Unlike the other structures, the line-of-sight velocity dispersion for this cluster includes two distinct components}\\
}
\end{table*}

Given the broad redshift range spanned by our sample, these structures are likely in very different evolutionary states, from non-virialized protoclusters at $z > 2$ \citep[e.g.,][]{Overzier2016,Alberts2022}, to more collapsed, quasi-virialized systems at lower redshifts.  As such, measurements of velocity dispersion can be vital for characterizing galaxy clusters, allowing us to probe both the internal dynamics of a cluster and the dynamics of sub-structures relative to one another (and potentially estimate cluster mass depending on whether a system is virialized).  Given the relatively small samples of galaxies, we measure the line-of-sight velocity dispersions for each structure using the Gapper method with jackknifed confidence intervals \citep{Beers1990}.  As seen in Table~\ref{tb:clusters}, our derived velocity dispersions range from 168\,km\,s$^{-1}$ to 833\,km\,s$^{-1}$.  For the double-peaked system at $z$ = 1.87, the global velocity dispersion reflects the complex sub-structure rather than a relaxed virialized halo.

\subsubsection{Characterizing Cluster/Protocluster Galaxies}\label{sect:Results-ClustersGal}

\begin{figure*}
\begin{center}
\includegraphics[scale=0.35,trim={0.4in 0.15in 0.in 1.7in},clip=true]{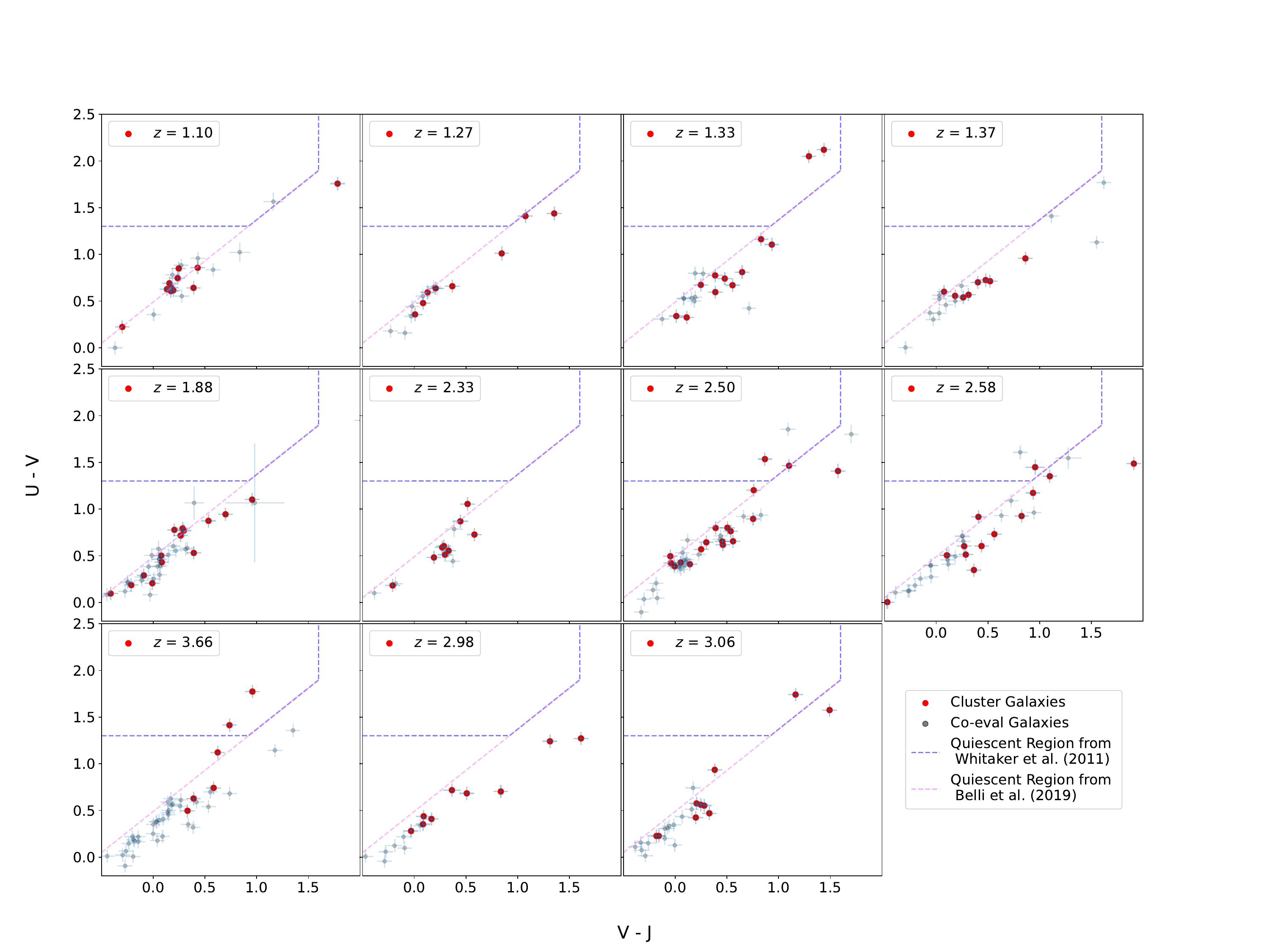}

\caption{Color-color diagrams for each structure in our sample.  Each plot shows Cigale modeled rest-frame colors. Red points correspond to galaxies in structures as defined by our density measures (see Section~\ref{sect:Clusters}), while the gray points are again co-eval galaxies that we do not classify as being cluster members either because of their density or because they are below our magnitude limits.  The blue dashed region in the upper left corner corresponds to the selection of quiescent galaxies following \citet{Whitaker2011}, while the pink diagonal dashed line corresponds to the criteria presented in \cite{Belli2019}.}

\label{Fig:Colorcolor}
\end{center}
\end{figure*}

Given that our sample spans 1.0 $\leq z \leq$ 3.1, the constituent galaxy populations are expected to reflect distinct evolutionary stages.  At $z$ $\sim$ 1, cluster environments are typically dominated by prominent red sequence populations, featuring large fractions of quenched galaxies across a wide range of stellar masses \citep[e.g.,][]{Gladders2000,Stott2009,Mei2009,Muzzin2009,Rudnick2012,Cooke2015,Cerulo2016,Lee-Brown2017,Tomczak2019,Lemaux2019,Golden-Marx2019}.  In contrast,  protoclusters at $z$ $>$ 2, are generally dominated by actively star-forming galaxies \citep[e.g.,][]{Casey2015,Hill2020,Shi2021,Staab2024,Edward2024,Forrest2024,Sikorski2026,Golden-Marx2026}, highlighting these pre-virialized structures as dynamic environments for rapid stellar mass assembly, although notable counterexamples hosting elevated quiescent fractions exist \citep[e.g.,][]{Willis2020,McConachie2022,McConachie2025,Toni2025b}.

To characterize these populations, we plot magnification corrected, rest-frame UVJ color-color diagrams for each structure (see Fig.~\ref{Fig:Colorcolor} and Section~\ref{Sect:SED} for how we estimated rest-frame colors).  Adopting the conservative $UVJ$ selection criteria from \citet{Whitaker2011} (see the blue region in Fig.~\ref{Fig:Colorcolor}), we identified eight quiescent galaxies, with the protoclusters at $z$ = 2.50 and $z$ = 2.66 each hosting multiple quiescent galaxies.  Expanding this selection using the criteria from \citet{Belli2019} (see the pink line in Fig.~\ref{Fig:Colorcolor}) increased the quiescent sample to 26 galaxies.  For the remainder of the paper, any quiescent galaxies discussed fall within the extended region from \citet{Belli2019}.   

\begin{figure*}
\begin{center}
\includegraphics[scale=0.35,trim={0.4in 0.15in 0.0in 1.7in},clip=true]{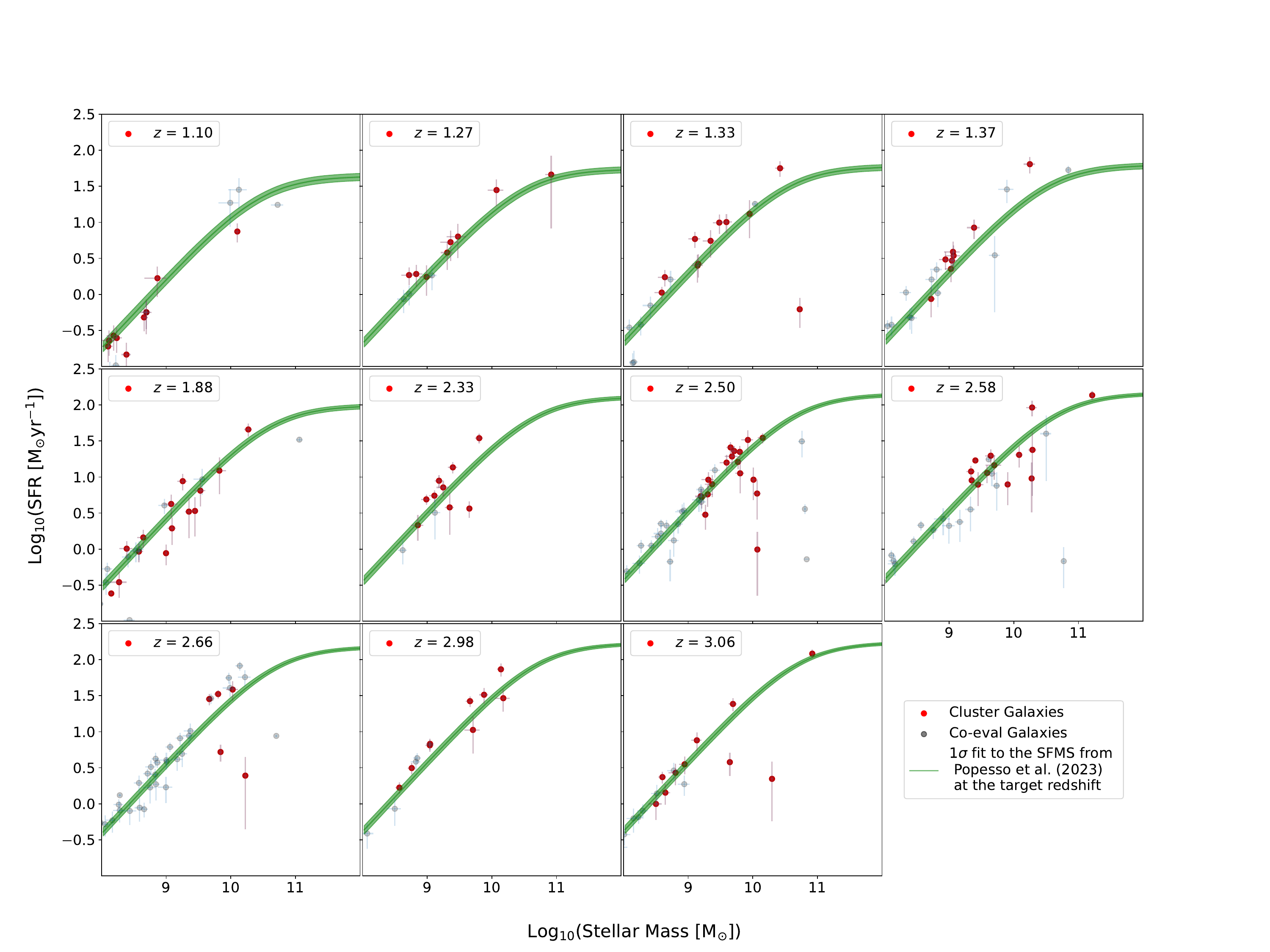}

\caption{Star-forming main sequence plots for the 11 structure candidates in this sample.  The green lines correspond to estimates of the star-forming main sequence from \citet{Popesso2023}.  As in Fig.~\ref{Fig:Colorcolor}, the red points correspond to galaxies in structures as defined by our density measurements (see Section~\ref{sect:Clusters}), while the gray points are again co-eval galaxies that we do not classify as being cluster members either because of their density or because they are below our magnitude limits.}

\label{Fig:MassSFR}
\end{center}
\end{figure*}

To further assess star-forming activity, we plot star formation rate as a function of stellar mass and compare our sample to the star-forming main sequence (SFMS) parametrization from \citet{Popesso2023}.  As shown in Fig.~\ref{Fig:MassSFR}, our magnification corrected values exhibit strong agreement within the 1$\sigma$ scatter of the star formation rate main sequence across the entire cluster sample.  At $z$ $>$ 2, we find little evidence for widespread quenching among lower mass  galaxies (M$_{*}$ < 10$^{9.5}$ M$_{\odot}$) in either cluster (red) or non-cluster environments (gray), although this regime approaches our photometric completeness limits.   

\subsubsection{Correcting for Multiple-Imaged Galaxies}\label{sect:DoubleImages}
Because we are characterizing structures in a strong gravitational lensing field, these structures can potentially include multiply imaged galaxies.  To prevent double-counting, we used our lens model (Section~\ref{sect:Lensing}) to predict the positions of potential counter-images for all spectroscopic targets within our candidate cluster redshift slices.  Cross-matching these predicted positions with our spectroscopic catalog revealed counter-images for three targets: one system at $z = 1.87$ with four counter-images, and two systems at $z = 2.66$ and $z = 3.05$, each with two counter-images. These multiple images correspond to previously identified systems (systems 2, 7, and 26 in \citealp{Bergamini2023a}), which are included as observational constraints in the strong lensing model. We removed these additional images from our catalog and re-ran the structure detection algorithm. In practice, however, after accounting for lensing magnifications, all identified counter-images fell below our strict flux completeness thresholds (Table~\ref{tb:clusters}) and would not have been selected as individual cluster members.

\subsection{Identifying Galaxy Clumps} \label{sect:Clump-results}

\begin{figure}
\begin{center}
\includegraphics[scale=0.55,trim={0.0in 0.in 0.0in 0.5in},clip=true]{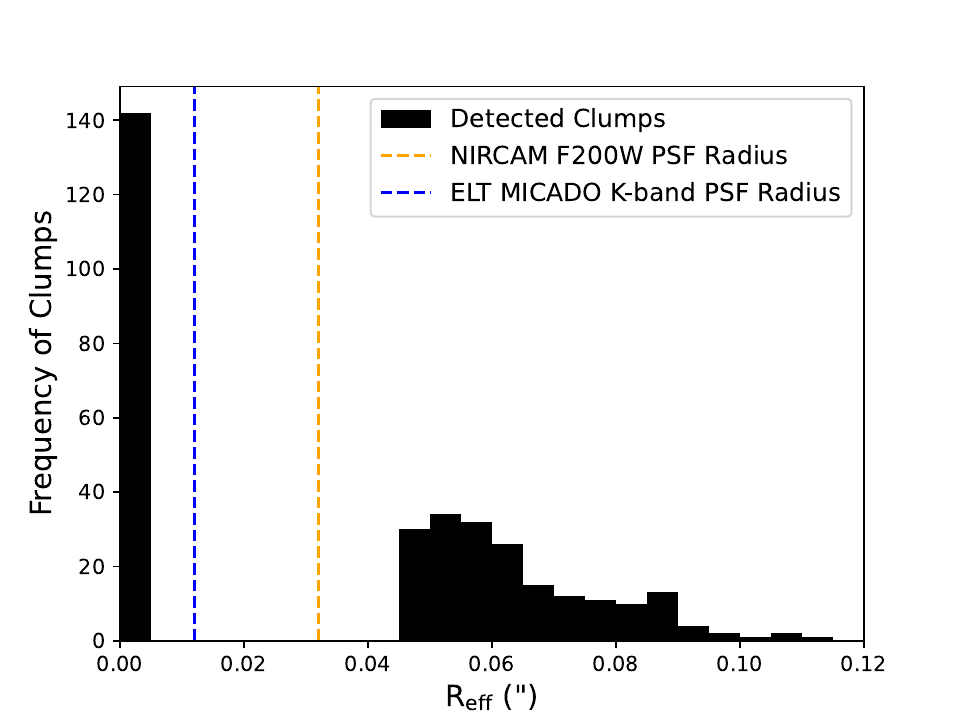} 

\caption{Detected radius of clumps in arcseconds.  These are the measured effective radii (not lens-corrected) of each bright clump estimated by transforming the SEP measured semi-major and minor axes and accounting for PSF.  We modeled clumps as point sources if they were smaller than 1.5\,r$_{\mathrm{F200W-PSF}}$.  We find $\sim$ 42.2$\%$ of our clumps are not resolved.  The radius of the F200W PSF (what are images are smoothed to) is shown by the dashed orange line and the radius of ELT MICADO in K-band is shown by the blue dashed line. Although the majority of clumps are larger than an individual pixel (0$\farcs$04 = 1 pixel), these clumps are still much smaller than the typical aperture radius used to measure galaxy fluxes.}

\label{Fig:Clumps-Radius}
\end{center}
\end{figure}

Given that most of our cluster galaxies are actively star-forming (Section~\ref{sect:Results-ClustersGal}), we expected a high prevalence of clumpy structures, in line with studies at cosmic noon \citep[e.g.,][]{Shibuya2016,Guo2016,Kalita2025,Sok2026,delaVega2026,Zhu2026,Puskas2026} and simulations showing clumpy galaxies are predominantly starbursting \citep[e.g.,][]{Wang2026}.  As described in Section~\ref{sect:Clump-Dection}, we detect 338 clumps across 110 galaxies, with sizes ranging from unresolved sources (smaller than the 1.5 $\times$ the half-width-half max of the PSF) to $\sim$ 0$\farcs$12 (see Fig.~\ref{Fig:Clumps-Radius}).  Despite our sample including moderately lensed galaxies, $\sim$42$\%$ of bright clumps are non-resolved point sources and $\sim$ 70$\%$ of all clumps are non-resolved.  

\begin{figure}
\begin{center}
\includegraphics[scale=0.55,trim={0.0in 0.in 0.0in 0.5in},clip=true]{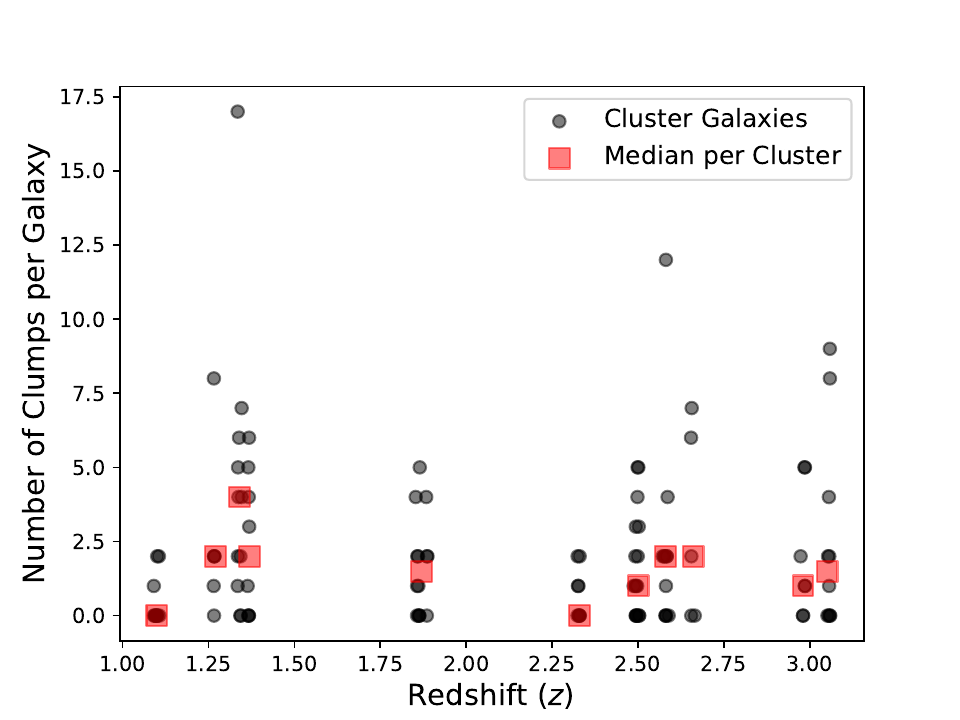}
\caption{Number of clumps as a function of redshift. For each galaxy in a structure, we plot the number of clumps (after removing potential proto-bulges) brighter than the clump magnitude limit reported in Table~\ref{tb:clusters}.  The median number of clumps per galaxy in each structure is shown in red.}
\label{Fig:Clumps-Multiplicity}
\end{center}
\end{figure}

Overall, the majority of our galaxies host multiple star-forming clumps.  Because our 11 clusters span discrete redshift windows across $1.0 \le z \le 3.1$, we evaluated potential redshift evolution in the clump multiplicity, the number of detected clumps per galaxy (see Fig.~\ref{Fig:Clumps-Multiplicity}).  Except for the structures at $z =$ 1.10 and $z =$ 2.32, we found relative uniformity in the median clump multiplicity.  However, because our detection thresholds are defined using apparent magnitude limits, rather than a fixed rest-frame absolute magnitude limit or stellar mass cut, and because host galaxy populations and their magnifications vary across structures, we cannot fully disentangle physical evolution from depth-dependent selection effects toward lower redshifts. 
Interestingly, the structure at $z$ = 1.10 lacks clumpy galaxies ($N_{clumps}$ $>$ 2) and has a median clump count and clump luminosity fraction of zero (see the left panel of Figure~\ref{Fig:Clumps-Multiplicity-v2}).  As shown in  Fig.~\ref{Fig:Colorcolor} and Fig.~\ref{Fig:MassSFR}, the member galaxies in this structure, while not fully quenched, lie systematically below the star formation main sequence, which is different from the majority of the other galaxies in our sample.  However, using a KS-test, we compare both the number of clumps and the distribution of clump luminosity fractions in the $z =$ 1.10 structure to our entire sample and do not find evidence that the distribution of clumpy galaxies is statistically distinct from our overlying sample ($p$ = 0.089). Thus, despite the apparent difference, these clumps are most likely drawn from the same sample distribution.  Although previous results show a declining fraction of clumpy galaxies at $z$ $<$ 1.0 \citep[e.g.,][]{Shibuya2016}, we lack statistical evidence of this in our sample and likely need more robust accounting of the truly quiescent, and likely non-clumpy galaxies, to show this evolution. 

\subsection{Global properties of clumpy galaxies}

\begin{figure*}
\begin{center}
\includegraphics[scale=0.4,trim={0.0in 0.in 0.5in 0.45in},clip=true]{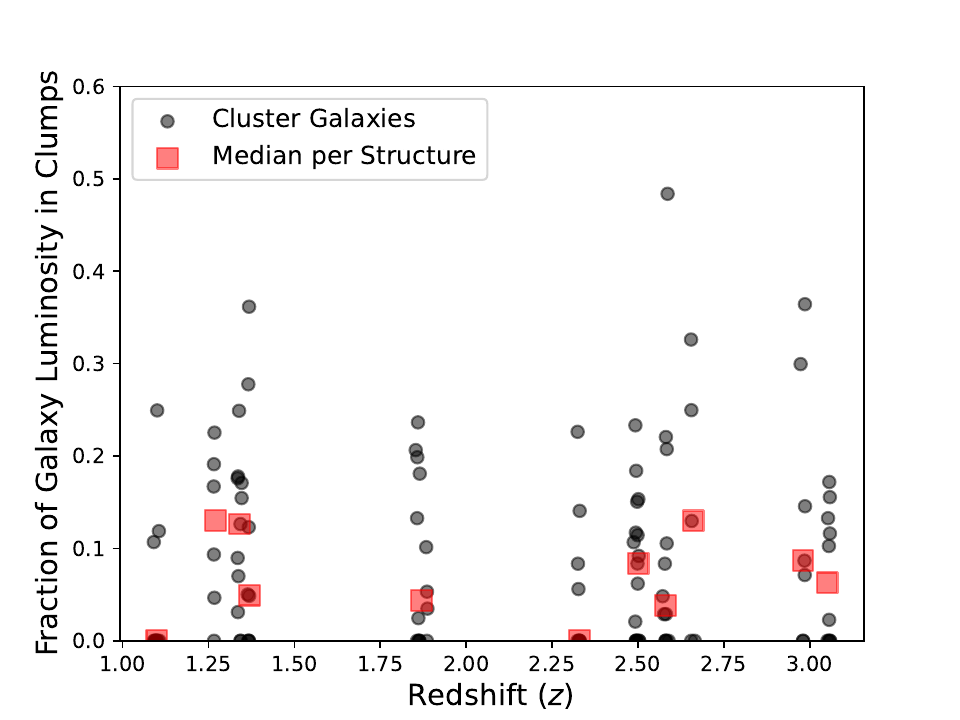}
\includegraphics[scale=0.4,trim={0.0in 0.in 0.5in 0.5in},clip=true]{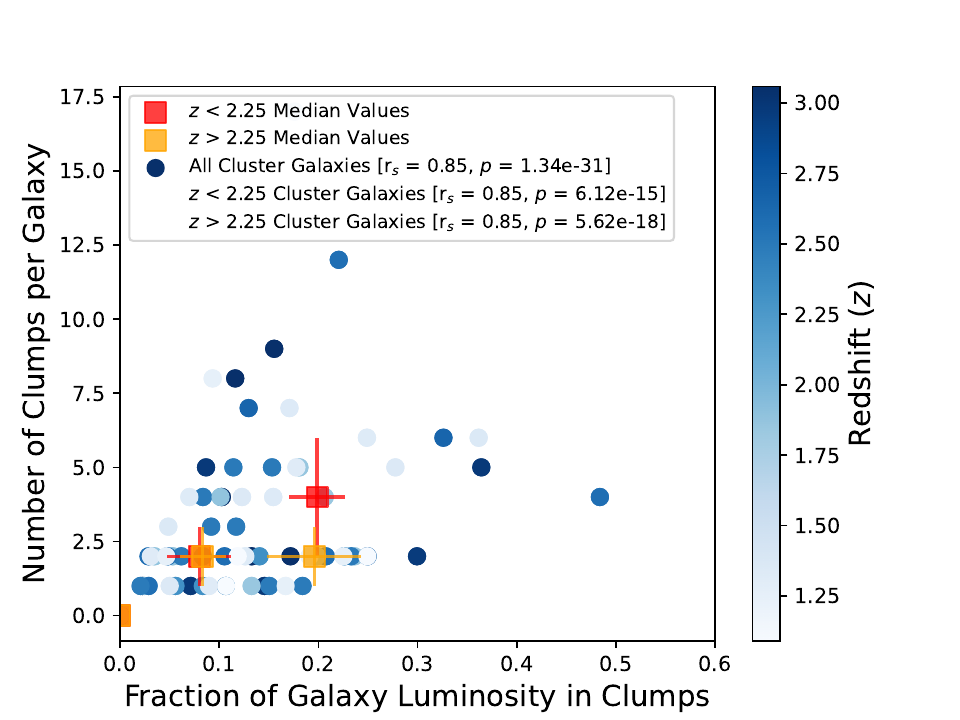}
\includegraphics[scale=0.4,trim={0.0in 0.in 0.5in 0.5in},clip=true]{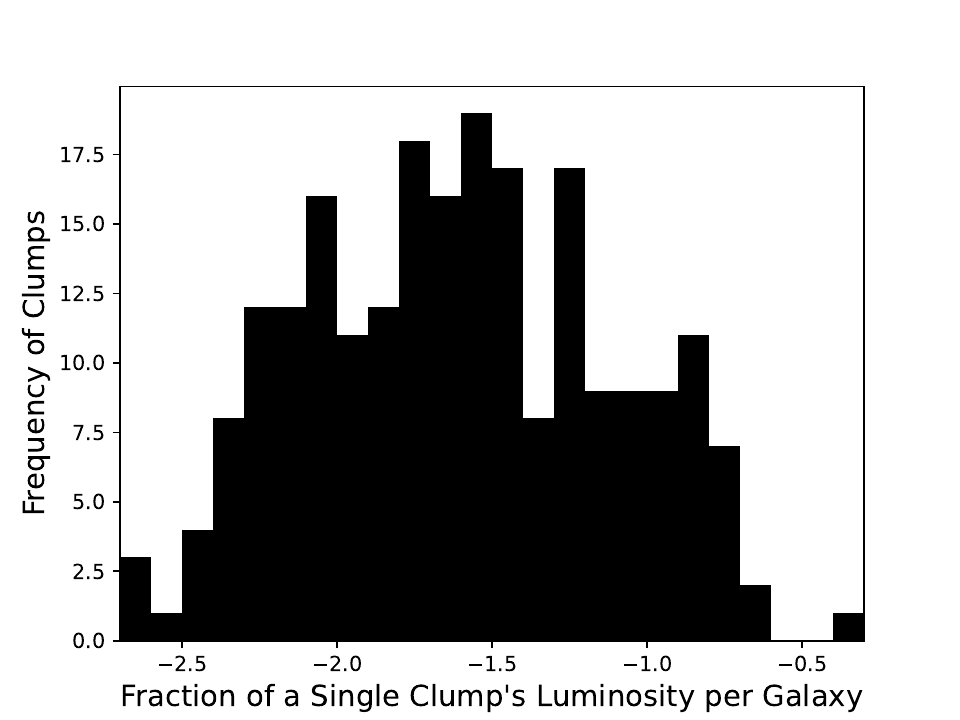}

\caption{Comparison of the fraction of rest-frame U-band light in clumps. Left: The fraction of luminosity as a function of redshift. Similar to Figure~\ref{Fig:Clumps-Multiplicity}, the red squares show the median fraction of the luminosity in clumps and we do not include the fraction of luminosity from proto-bulges or clumps fainter than our magnitude limits.  Middle: The number of clumps as a function of the fraction of the host galaxy's luminosity encased in clumps.  The color of each point reflects the redshift of the galaxy.  We find 42 of our galaxies ($\sim$ 38.2$\%$) have no clumps and that this fraction is constant between our lower and higher redshift bins.  The red squares show the median values among the lower redshift sample ($z <$ 2.25), while the orange squares show the median values among the higher redshift sample ($z >$ 2.25).  The error bars in each reflect the sample spread. Right: Histogram of the fraction of rest-frame U-band luminosity encased in clumps.}

\label{Fig:Clumps-Multiplicity-v2}
\end{center}
\end{figure*}

Owing to the broad diversity in individual clumps sizes and fluxes, the range of intrinsic luminosities associated with each clump differs dramatically (see the right panel of Figure~\ref{Fig:Clumps-Multiplicity-v2}).  Furthermore, despite our magnitude limits, Malmquist biasing means that we are more likely to detect brighter clumps, making it difficult to compare the number of clumps across galaxies of differing brightness. To better account for this difference and present a more physically motivated parameter, we instead parameterize the number of clumps as the summed fraction of the host galaxy's luminosity found in clumps (e.g., the summed rest-frame U-band clump luminosity divided by the total galaxy's rest frame U-band luminosity including clumps).  As shown in the lefthand panel of Figure~\ref{Fig:Clumps-Multiplicity-v2}, we find a similar fraction of luminosity encased in clumps across all our clusters, with the medians in the clusters at $z$ = 1.10 and $z$ = 2.35 again being zero.  Additionally, as shown in the middle panel of Figure~\ref{Fig:Clumps-Multiplicity-v2}, we explore the relationship between the number of clumps and the fraction of luminosity in clumps.  Following \citet{Golden-Marx2021} and the statistical framework of \citet{Fowler2009}, we quantify these trends using Spearman rank correlation tests, where r$_{s}$ is the strength of the correlation and $p$ is the ability to reject the null hypothesis.  We find a statistically strong correlation between the fraction of luminosity in clumps and the number of clumps (r$_{s}$ = 0.85, $p$ = 1.34$\times$10$^{-31}$).  Although there is a wide spread in the number of clumps that corresponds to $>$ 20$\%$ of luminosity in clumps, the strength of this trend remains high among both the lower redshift portion of our sample ($z <$ 2.25; r$_{s}$ = 0.85, $p$ = 6.12$\times$10$^{18}$) and the higher redshift portion of our sample ($z >$ 2.25; r$_{s}$ = 0.85, $p$ = 5.62$\times$10$^{-18}$).

\begin{figure*}
\begin{center}
\includegraphics[scale=0.315,trim={0.3in 0.1in 0.5in 0.85in},clip=true]{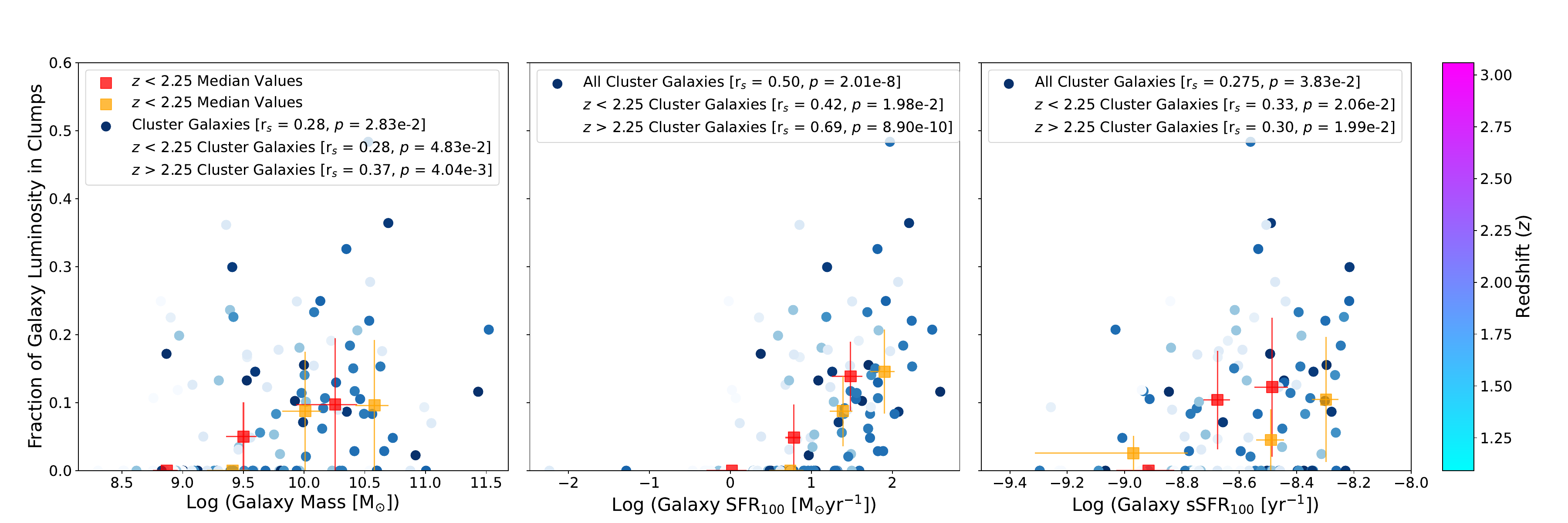} 
\caption{Measurements of the luminosity enclosed in clumps compared to galaxy properties. Left: The fraction of luminosity enclosed in clumps as a function of galaxy mass. Middle: The fraction of luminosity enclosed in clumps as a function of galaxy SFR. Right: The fraction of luminosity enclosed in clumps as a function of sSFR.  In each plot, the color of the points reflect the host galaxy's redshift, while the red and orange squares show binned median values in the lower ($z <$ 2.25; red) and higher ($z >$ 2.25; orange) redshift samples.  These red and orange points are estimates of the median of equal sized galaxy bins.  We quantify the strength of each trend using a Spearman test and include the strength and likelihood to reject the null hypothesis in the legend of each plot. }

\label{Fig:Clumps-2}
\end{center}
\end{figure*}

To contextualize our results, we compared our CIGALE derived physical parameters (stellar mass, SFR, sSFR) to the fraction of the luminosity encased in clumps. As shown in Fig.~\ref{Fig:Clumps-2}, we find that more star-forming galaxies systematically host a higher number of clumps.  We find a statistically significant moderate positive correlation between fraction of luminosity in clumps and SFR (r$_{s}$ = 0.50, $p$ = 2.01$\times$10$^{-8}$).  We further examine this trend with SFR by dividing our sample into two redshift subsamples, $z <$ 2.25 and $z >$ 2.25.  Although we find statistically moderate correlations with strong evidence to reject the null hypothesis in each subsample ($z < 2.25$ - r$_{s}$ = 0.42, $p$ = 1.98$\times$10$^{-2}$; $z >$ 2.25 - r$_{s}$ = 0.69, $p$ = 8.90$\times$10$^{-10}$), the trend is stronger among the higher redshift sample.  The stronger dependence on total SFR, suggests that clumpy star formation is a primary driver of stellar mass growth in these galaxies, particularly among higher redshift protocluster galaxies. 

In contrast, we find weaker, though still statistically robust trends between the fraction of luminosity in clumps and stellar mass and sSFR in clumps (r$_{s}$ = 0.28, $p$ = 2.83$\times$10$^{-2}$ and r$_{s}$ = 0.28, $p$ = 3.83$\times$10$^{-2}$), implying a slight trend between galaxy mass and star formation efficiency and the fraction of luminosity in clumps.  Removing the small sample of quiescent galaxies (Section~\ref{sect:quiescent}) does not significantly alter the trend with stellar mass.  Although not shown, we did examine these trends with the number of clumps instead of the fraction of luminosity in clumps and found similar results to those reported, although we found no statistical evidence for the trend with sSFR when using the number of clumps.

\section{Discussion}\label{sect:Discussion}

\subsection{Contextualizing our structure sample}
Although our sample spans a wide range in redshift, line-of-sight velocity dispersions provide a valuable diagnostic for understanding the evolutionary state of clusters.  As shown in Table~\ref{tb:clusters}, most of our structures have velocity dispersions below 400\,km\,s$^{-1}$.  Among systems at $z <$ 1.5, only one has a velocity dispersion above 700\,km\,s$^{-1}$.  For context, \citet{Lemaux2019} reported velocity dispersions for 56 clusters and groups in the ORELSE survey at 0.5 $< z <$ 1.4, and defined an empirical upper limit of $\sigma_v \approx 550\,\mathrm{km\,s^{-1}}$ for galaxy groups.  Similarly, \citet{Biviano2021} analyzed 14 clusters from the GOGREEN sample at 0.8 $< z < $ 1.4 and reported velocity dispersions ranging from 400\,km\,s$^{-1}$ to 1200\,km\,s$^{-1}$.  Based on these benchmarks, two of our structures at $z$ $<$ 1.5 are better classified as galaxy groups, and only the cluster at $z$ = 1.10 ($\sigma_{\parallel}$ = 832.5$\pm$162.2\,km\,s$^{-1}$) qualifies as a fully collapsed cluster halo.  Importantly, \citet{Biviano2021} found that the velocity dispersion decreased slightly beyond 0.3\,r$_{200}$, so even if our structures are better characterized as cluster cores, these velocity dispersions likely represent upper limits.  Additionally, \citet{CWatson2025} reported a velocity dispersion of $700\,\mathrm{km\,s^{-1}}$ and a virial mass of $2.2 \times 10^{14}\,\mathrm{M}_\odot$ for a $z = 1.81$ cluster.  Because our structure at $z =$ 1.87 exhibits a higher velocity dispersion due to it encapsulating two nearby cluster peaks, it is difficult to directly compare these two clusters to determine if our $z =$ 1.87 structure might be a similarly massive cluster. 

In protoclusters ($z >$ 2), lower velocity dispersions indicate lower relative velocities between member galaxies, which can enhance galaxy-galaxy merger rates \citep[e.g.,][]{Cucciati2014,Toshikawa2020,Lemaux2022,Liu2024, Giddings2025}.  In particular, the meta-analysis by \citet{Liu2024} across 2 $<$ $z$ $<$ 3.1 demonstrated that lower velocity dispersions correlated with higher close-pair fractions and higher merger rates. Although gravitational lensing distortions preclude a standard close-pair analysis in this field, all six of our protoclusters at 2 $<$ $z$ $<$ 3.1 have velocity dispersions below 383\,km\,s$^{-1}$, which could indicate significant merger activity.  Combined with the local galaxy overdensities identified in our core regions, these low velocity dispersions point to an environment favorable for galaxy-galaxy interactions. As suggested by \citet{Lemaux2022}, enhanced merger activity in high-$z$ protoclusters may also drive the turnover in the star formation rate-density relation seen in the early universe compared to low-$z$ clusters.  As such, if a fraction of our detected clumps form via merger-driven gas compression \citep[e.g.,][]{Ribeiro2017} or are in fact multi-component galaxy-galaxy mergers, the high prevalence of clumpy star-forming galaxies in our sample (Section~\ref{sect:Clump-Dection}) provides observational support for frequent interaction events in early structure cores.

\subsection{Populations of Cluster Galaxies}
\subsubsection{Quiescent Populations}\label{sect:quiescent}

As seen in Fig.~\ref{Fig:Colorcolor}, our four lower redshift structures ($z < 1.5$) host only seven quenched galaxies ($\sim$ 18$\%$) according to the criteria of \citet{Belli2019} (and only two following the criteria of \citealp{Whitaker2011}).  Almost all of these galaxies are at $z =$ 1.10.  Assuming our spectroscopic sample adequately traces the member galaxy population, this quiescent fraction, particularly among massive galaxies ($\sim$ 17$\%$) is below the 20 - 30$\%$ of massive galaxies (M $>$ 10$^{10.8}$\,M$_{\odot}$) and 20$\%$ of low-mass galaxies (10$^{10}$\,M$_{\odot}$ $<$ M $<$ 10$^{10.8}$\,M$_{\odot}$ at 1.0 $< z <$ 1.5 in \citet{Lemaux2019}, suggesting that we are probing dynamically young galaxy structures, similar to the cluster at $z$ = 1.8 identified in \citet{Rinaldi2026}.  Furthermore, the small quiescent fractions in our two galaxy group candidates ($z <$ 1.5; $\sigma_{\parallel} <$ 200\,km\,s$^{-1}$) strengthens the hypothesis that these systems are high-$z$ galaxy groups, since groups typically have lower quenched fractions at all redshifts relative to their massive counterparts \citep[e.g.,][]{Wetzel2012,Toni2025b}.

Among our lower redshift structures, only the system at $z$ = 1.10 hosts a notable population of quenched galaxies.  As seen in Fig.~\ref{Fig:Colorcolor} and Fig.~\ref{Fig:MassSFR}, this cluster contains a dominant massive galaxy, which, although not quiescent, lies below the $z = 1.10$ star-forming main sequence \citep{Popesso2023}, alongside several lower-mass quenching candidates seen above the dashed line in Fig.~\ref{Fig:Colorcolor} showing the trend from \citet{Belli2019}.  Interestingly, this cluster also lacks highly clumpy galaxies (see Fig.~\ref{Fig:Clumps-Multiplicity}).  Given that cluster starburst activity is often driven by mergers \citep[e.g.,][]{Brodwin2013} and in simulations, starbursting galaxies are clumpier \citep{Wang2026}, the combined prevalence of quenching targets and the absence of clumpy morphologies likely mark the onset of widespread environmental quenching in this structure.

Because our analysis consists exclusively of spectroscopically-confirmed galaxies, selection effects may bias our derived passive fractions, particularly at lower redshift, where we  detect fewer massive quiescent galaxies than in our structures at 1.8 $< z < $ 3.1  Because our secure spectroscopic catalog (see Section~\ref{sect:data}) requires galaxies to have multiple emission lines or spectral continuum features for certain surveys, passive galaxies with weak line emission are intrinsically harder to confirm spectroscopically.  As shown in Appendix~\ref{Appendix}, only $\sim$ 30$\%$ of red, potentially quiescent galaxies are spectroscopically confirmed, compared to the closer to 50$\%$ of bluer galaxies.  Additionally, the wavelength coverage of NIRSpec and NIRCam WFSS is also not ideal for detecting these galaxies at $z <$ 1.5.  Examining photometric redshift catalogs (see Appendix~\ref{Appendix}), reveals an elevated quiescent fractions across all structures relative to the spectroscopic sample.  However, given the narrow line-of-sight velocity bounds of cluster cores and the complications of gravitational lensing, assigning photometric targets to individual structures carries substantial uncertainty.  Consequently, our measured quiescent fractions should be interpreted as lower limits on the true passive population. This limitation is exemplified by the $z = 2.58$ protocluster, where \citet{Pan2025} identified a substantial population of lower-mass quiescent members using combined photo-$z$ and spec-$z$ selections (Section~\ref{sect:cluster-comp-lit}), many of which lack spectroscopic confirmation in our catalog.

Despite the overall scarcity of quiescent galaxies, four of the five highest redshift protoclusters host at least one massive quiescent galaxy (M $>$ 10$^{10}$\,M$_{\odot}$). These systems may resemble the quiescent proto-BCG identified in BOSS1244 ($z$ = 2.24; \citealp[e.g.,][]{Shi2024}). In particular, the massive quiescent galaxy in our $z = 2.66$ protocluster, also discussed in \citet{Marchesini2023} and \citet{Bevacqua2026}, represents one well-studied quenched structure member in our sample.

\subsubsection{Populations of Star-Forming Galaxies}

Interestingly, despite the relative scarcity of quiescent populations, none of the protoclusters display the strong starbursting cores reported in some high-$z$ systems \citep{Shimakawa2018a,Shimakawa2018}.  Instead, as shown in  Fig.~\ref{Fig:MassSFR}, the member populations align remarkably well with the empirical star-forming main sequence.  The absence of extreme starburst activity may partly stem from selection effects in our rest-frame optical spectroscopy, which lacks sensitivity to strong Lyman-$\alpha$ emitters that can populate protocluster cores \citep[e.g.,][]{Cai2017,Umehata2025}.

Nevertheless, agreement with the star-forming main sequence alongside a low quiescent fraction is also predicted by numerical simulations of protoclusters \citep[e.g.,][]{Muldrew2018,Remus2023}, and mirrored in observational studies at similar redshifts \citep[e.g.,][]{Forrest2024,Rinaldi2026,Bertemes2026}.  Specifically, \citet{Forrest2024} noted the lack of an elevated quiescent population in the Elentari protocluster at $z$ = 3.3 when compared to a co-eval field.  Although we lack a robustly characterized field sample, both Fig.~\ref{Fig:Colorcolor} and Fig.~\ref{Fig:MassSFR} show comparable passive fractions in both cluster and field environments,  suggesting that the few massive quiescent galaxies in our structures likely quenched via intrinsic mass-dependent processes rather than environmental mechanisms.

\subsection{Comparing our Structure Detections to Previously Identified Structures in the Abell 2744 Field}\label{sect:cluster-comp-lit}
As the field surrounding Abell 2744 has been well-studied with JWST, comparing our candidate structures with previous literature provides essential context.  Eight of our structures have been reported in prior studies, while three (at $z$ = 1.27, 2.33, and 2.95) represent newly identified systems.  Five of these structures ($z$ = 1.10, $z$ = 1.33, $z$ = 1.36, $z$ = 1.87, $z$ = 2.66) were identified using NIRISS observations in \citet{Watson2025}, which treated the systems at $z$ = 1.33 and $z$ = 1.36 as single structure.  Because our spectroscopic compilation included the NIRISS catalogs from \citet{Watson2025}, we found strong agreement in terms of cluster memberships.  Any discrepancies primarily stem from our inclusion of additional spec-$z$ catalogs and our brightness/density thresholds, which exclude some fainter \citet{Watson2025} targets.  Unsurprisingly, the interacting pair discussed in \citet{Watson2025} and identified in \citet{Wang2020} corresponds to one of the few spectroscopic pairs of galaxies that we treated as a single system in our clump detection to best capture the clumps belonging to the entire interacting system and not just one half of it (see Section~\ref{sect:Clump-results}).  

Similarly, \citet{Naidu2024} identified prominent overdensities at $z$ = 2.50 and $z$ = 2.58, along with a candidate structure at $z$ = 1.87, all of which match peaks in our sample.  However, the spatial scales evaluated in these two studies do differ: \citet{Naidu2024} probed environments over  5\,cMpc aperture, whereas we employ a 1\,Mpc (3.5\,cMpc scale at $z = $2.5) search radius.  This distinction may explain why \citet{Naidu2024} found that many of the most massive galaxies in their sample reside in overdensities, whereas our $1\,\mathrm{Mpc}$ core selection shows that these massive hosts do not always coincide with central density peaks. Importantly, \citet{Naidu2024} also included an abundance of low mass galaxies (M $<$ 10$^{7}$\,M$_{\odot}$), well below our detection threshold.  Thus, it is possible that the massive galaxies discussed in \cite{Naidu2024} may reside within the extended protocluster environment, rather than inside our central peaks/overdensities.  Such scale-dependent offsets align with cosmological simulations: \citet{Baxter2026} modeled the evolution of massive JWST-selected galaxies, finding that large-scale structure is critical to their evolution and that the most massive $z >$ 2 galaxies are typically not the direct progenitors of the most massive $z =$ 0 cluster galaxies.  Consequently, the fact that the most massive galaxies at a given epoch are not identified as central cluster members at $z > 2$ in this work does not preclude our massive cluster galaxies from evolving into massive low-$z$ BCGs, particularly through merger-driven growth, as suggested by their clumpy morphologies. 

Other studies have identified additional candidate members that lack spectroscopic confirmation in our spec-$z$ catalog and are therefore excluded from our analysis. For instance, \citet{Whitaker2025} identified a clumpy massive quiescent galaxy at $z$ $\sim$ 2.53 associated with the broad structure at $z$ = 2.5 \citep[e.g.,][]{Naidu2024, Pan2025}, along with two other photometric quiescent galaxies at 2.4 $< z <$ 2.5. 
Similarly, \citet{Pan2025} identified a protocluster at 2.575 (2.5 $<$ $z$ $<$ 2.65).  While their redshift selection overlaps significantly with our $z =$ 2.58 protocluster (notably \citealp{Pan2025} did not incorporate the NIRISS catalogs from \citealp{Watson2025} or the GLASS-JWST spectroscopy from \citealp{Mascia2024}) their broader redshift range also encompasses members from our distinct $z$ = 2.50 and $z$ = 2.66 protoclusters.  A direct target-by-target comparison is limited by the absence of a published membership catalog in \citet{Pan2025}; nevertheless, comparing the rest-frame $UVJ$ color-color diagrams for the clusters at $z$ = 2.50, 2.58, and 2.66 with their Fig. 7 reveal consistent color distributions.  However, likely because of our magnitude cut, we cannot characterize the low-mass quiescent population reported in this protocluster.  Additionally, \citet{Pan2025} intentionally excluded $z$ $\sim$ 2.4 and 3.1 from their background field selection to avoid structural contamination— redshift intervals, where we independently confirm three overdense candidates.

Incorporating the photometric quiescent candidates from \citet{Whitaker2025} and \citet{Pan2025} sheds further light on the environmental quenching discussion in Section~\ref{sect:quiescent}. Given that protoclusters hosting these galaxies are the richest and most extended in our sample, it is possible that the massive galaxy in \citet{Whitaker2025} traces a secondary quiescent peak within this structure.  Given that this structure at $z =$ 2.50 already includes multiple massive quiescent galaxies, this would provide additional evidence that this cluster is unique among our sample and may show evidence of the build-up of an early forming red sequence, resembling mature $z > 2$ protoclusters such as those in \citet{Willis2020} or \citet{McConachie2022}, or the pair of massive quiescent galaxies, one of which is a proto-BCG, identified in the star-forming protocluster BOSS1244 at $z =$ 2.24 \citep{Shi2024}.  

Similarly, the protocluster at $z$ = 2.66 was also identified in \citet{Bevacqua2026}, who identified eight coeval members. Our core selection captures five spectroscopic members in our sample.  Apart from the massive quiescent galaxy reported by \citet{Marchesini2023}, the remaining targets in \citet{Bevacqua2026} either fall outside our density cuts or lack matching spec-$z$ confirmations in our catalog, suggesting that this system forms part of a larger, extended filamentary structure.

Lastly, the structure at $z$ = 3.05 was previously identified and studied in \citet{Wu2023} and \cite{Li2026}. \citet{Li2026} reported 23 total members (eight spectroscopic, including three within a $70\,\mathrm{kpc}$ core).  Our selection identifies ten spectroscopically confirmed members with redshifts ranging from 3.05 to 3.06. Notably, this structure hosts some of the most clumpy galaxies in our samples, including the grand design spiral identified in \citet{Wu2023}, which has nine bright clumps.  Importantly, \citet{Li2026} reported evidence of ram pressure stripping in a  spectroscopically-confirmed galaxy.  Although this target falls below our flux completeness threshold for cluster membership and is not included in our study of clumpy galaxies, its presence is another example of a clumpy, interacting galaxy within this structure. Targeted follow-up observations, of both the host galaxies and their constituent clumps will be essential to constrain clump ages, untangle the underlying dynamics, and reconstruct the star formation history of this active protocluster.

\subsection{Contextualizing the clumpy galaxies in the structures in the Abell 2744 field}

As shown in Fig.~\ref{Fig:Clumps-Multiplicity}, we identify 388 galaxy clumps across 110 cluster/protocluster galaxies.  While the majority of galaxies have at least one clumpy component, 42 of these galaxies (38.2$\%$) have no clumps and only host a proto-bulge.  As shown in Fig.~\ref{Fig:Clumps-2}, the majority of these galaxies without clumps are lower mass, lower star-forming rate galaxies.  However, we also identified a population of massive (M $>$ 10$^{10}$\,M$_{\odot}$) actively star-forming galaxies (SFR $>$ 10\,M${_{\odot}}$/yr) devoid of clumps.  These galaxies are particularly intriguing given the statistical correlations linking the fraction of luminosity in clumps to the total SFR.  While a fraction of these galaxies may represent bulge-dominant star-forming galaxies or active galactic nuclei, our conservative flux thresholds likely exclude a population of fainter or more compact clumps that require deeper observations to resolve. 

\subsubsection{A comparison to clump studies at cosmic noon}\label{sect:clump-lit}
Clumpy galaxy morphologies have been extensively documented in pre-JWST studies \citep[e.g.,][]{Wuyts2012,Shibuya2016,Guo2016,Sok2022}, strongly lensed fields \citep[e.g.,][]{Vanzella2021,Vanzella2022,Claeyssens2023,Claeyssens2025,Kalita2026-s1,Kalita2026-s2}, and recent high-resolution JWST imaging across cosmic noon \ \citep[e.g.,][]{Kalita2024,Kalita2025,Kalita2025b,Kalita2025c,Estrada-Carpenter2025,Sok2026a,Sok2026,delaVega2026,Zhu2026,Puskas2026}. However, a standardized definition for classifying "clumpy" galaxies remains lacking in the literature.  While several works \citep{Shibuya2016,Guo2016,Sok2022,Sok2026} define clumpy galaxies as those with more than 5$\%$ of their luminosity encased in clumps,  \citet{delaVega2026} require at least one prominent, off-axis clump (dependent on host galaxy stellar mass).  As shown in Fig.~\ref{Fig:Clumps-2}, when excluding central proto-bulges, 61.8$\%$ of our galaxies host at least one clump. Because virtually all of these galaxies also satisfy the L$_{\mathrm{clump}} > 0.05$\,L$_{\mathrm{galaxy}}$ criterion, adopting the presence of at least one non-bulge clump yields an overall clumpy fraction reported above. This result aligns well with clumpy fractions reported across similar redshifts in JWST field surveys, such as \citet{Sok2026} ($\sim$ 60$\%$ at $z \sim 2$ and $\sim 40\%$ at $z \sim 4$), \citet{delaVega2026} ($\sim$ 70$\%$ at $z =$ 2.75), and \citet{Puskas2026} ($\sim$ 55$\%$ at $z =$ 1 - 2). Given that \citet{Sok2026} provides the most comprehensive statistical baseline to date across $0.5 < z < 4$, a detailed comparison with their samples is essential to contextualize our findings.

Despite this broad numerical consistency, several systematic differences in sample selection and measurement techniques warrant detailed examination. First, at $z$ $>$ 2, our magnitude limits exclude most galaxies with masses M$_{*}$ $<$ 10$^{9.5}$\,M$_{\odot}$.  In contrast \citet{Sok2026} probed down to lower stellar masses, where they found that lower-mass galaxies host intrinsically feature fewer clumps (0 $<$ N$_{\mathrm{clumps}}$ < 4), particularly at 0.5 $<$ $z$ $<$ 3.  This mass dependence is partially reflected within our own sample: our structure at $z$ = 1.10 which hosts the highest fraction of low-mass galaxies, also contains the largest non-clumpy population.  However, this structure also hosts the most quenched galaxies, which should also depreciate clumpiness, making it difficult to disentangle the two in our analysis.  Second, our selection criteria are more inclusive: unlike \citet{Kalita2025b} and \citet{Sok2026}, we did not reject edge-on disk geometries, and by accounting for close neighboring components, our catalog retained major merger candidates. Third, on the algorithmic front, smoothed subtraction techniques (such as ours) tend to recover slightly higher clump flux fractions than wavelet decomposition algorithms \citep[][]{Sok2026}, consistent with our median clump fluxes being slightly brighter at $z < 3$ (Fig.~\ref{Fig:Clumps-Multiplicity-v2}, right panel) than their results.  These differences, along with any potential intrinsic environment differences, may account for discrepancies between our results.  

\begin{figure}
\begin{center}

\includegraphics[scale=0.55,trim={0.0in 0.0in 0.0in 0.5in},clip=true]{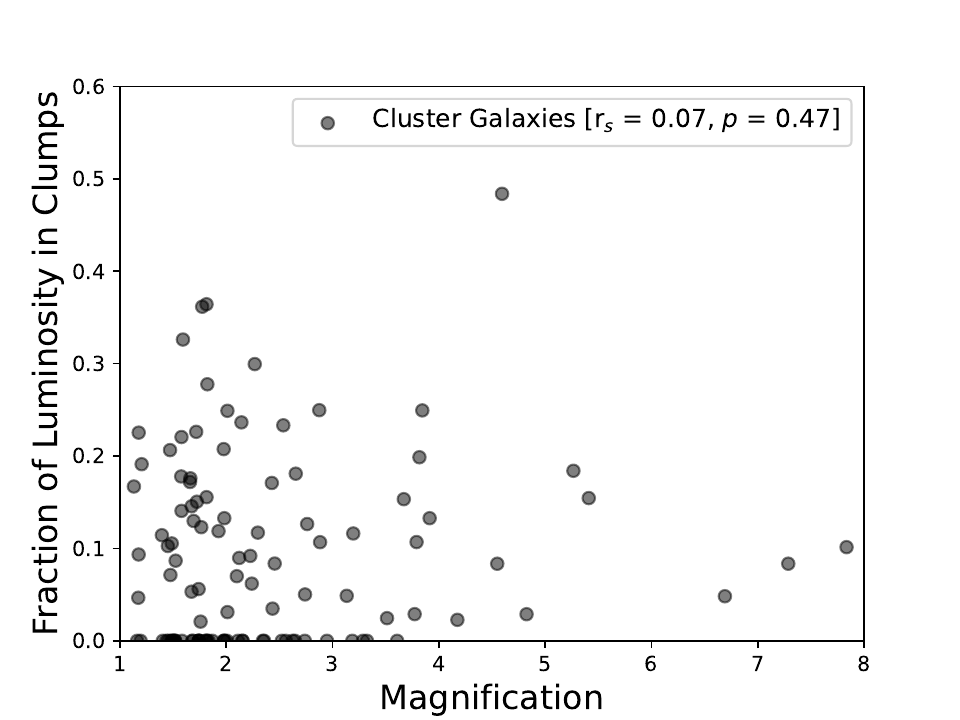} 

\caption{Fraction of galaxy luminosity in clumps as a function of magnification.  Although we have multiple extremely clumpy galaxies, we find no biases that strongly magnified galaxies have more clumps [r$_{s}$ = 0.07, $p$ = 0.47].}

\label{Fig:Clumps-magnification}
\end{center}
\end{figure}

Another notable distinction is our treatment of gravitational lensing. Our sample includes multiple magnified galaxies ($\mu$ $>$ 2; see Fig.~\ref{Fig:Clumps-magnification}), which were explicitly excluded by \citet{Sok2026}. We accounted for lensing effects by enforcing intrinsic, de-lensed magnitude completeness limits.  As shown in Fig.~\ref{Fig:Clumps-magnification}, we did not see any strong correlation between the fraction of luminosity in clumps and the reported magnification (r$_s$ = 0.07, $p$ = 0.47), confirming that lensing does not artificially distort our morphological statistics.  

Our results also align with dedicated lensed studies in Abell 2744 by \citet{Claeyssens2025}, who focused on more strongly-magnified galaxies (1.8 $<$ $\mu$ $<$ 300) and included higher redshift galaxies ($z <$ 5.5).  Despite different clump identification methods, their $z <$ 3 sample is dominated by galaxies hosting between 2 and 10 clumps, similar to our results.  Additionally, although they do not explicitly compare the number of detected clumps to the magnification, they report that the clumpiest galaxies (N$_{\mathrm{clumps}}$ $>$ 20) are primarily at $z$ $>$ 3, while the most magnified galaxies are at $z$ $\sim$ 4, which agrees with our results.  This behavior contrasts with the small sample of $z$ $<$ 6 strongly lensed clumpy galaxies in the field surrounding SMACS0723, where \citet{Claeyssens2023} found that extremely clumpy galaxies ($N_{\mathrm{clumps}} > 10$) were concentrated in highly magnified ($\mu > 7.9$) systems at $z > 2.2$, a regime not probed here, as the majority of our hosts feature moderate magnifications ($\mu < 2$). As a consequence, it is possible magnification can play a role depending on the cluster. 

Beyond individual clump detection, our key result is a robust positive correlation linking the fraction of rest-frame U-band luminosity in clumps to the total galaxy SFR, particularly among our protocluster sample.  A similar trend was recently reported by \citet{Zhu2026} across 3600 galaxies at 2 $<$ $z$ $<$ 8 from the JADES fields and in \cite{Puskas2026} across 10583 galaxies at 1 $< z <$ 3 in the JADES field. However, neither \citet{Sok2026} nor \citet{delaVega2026} reports trends between the star formation rate and the number of clumps, with \citet{Sok2026} observing a similar distribution of SFRs between their clumpiest and least clumpy galaxies.  

Interestingly, both \citet{Sok2026} and \citet{delaVega2026} found that clumpiness correlates primarily with host stellar mass.  This mass dependence is echoed in our sample with through our weak, statistically significant trend between the fraction of luminosity in clumps and the stellar mass.  However, it is possible that we find a much weaker trend due to our stricter magnitude limits limiting us to primarily massive galaxies.  Additionally, our weak, but statistically significant dependence on sSFR mirrors the HST COSMOS findings of \citet{Sok2022}, who observed a flat relation  between sSFR and the fraction of clumpy galaxies in a sample of 20,000 clumpy galaxies. This implies that while giant clumps trace total gas and mass accumulation, their formation does not necessarily boost global star formation efficiency.

\section{Conclusion}\label{sect:Conclusions}
In this work, we investigate the star-forming properties of clumpy galaxies within overdense environments at cosmic noon.  By combining multiple rich JWST spectrospic catalogs across the Abell 2744 field, we constructed a unified compilation comprising 2007 unique spectroscopically confirmed galaxies within the UNCOVER footprint, including 885 targets at $1.0 \leq z \leq 4.0$. This dataset allows us to characterize the environments of spectroscopically-confirmed galaxies at cosmic noon. The main results are summarized below.  

\begin{itemize}
    \item By measuring 3D overdensities and accounting for the lensing-corrected magnitude limits, we identified 11 cluster/protocluster structures at 1.0 $< z <$ 3.1, 8 of which were previously reported in spectroscopic or photometric studies.
    \item These structures, which may be better characterized as protocluster peaks at high redshifts ($z >$ 2.4), and galaxy groups at low redshifts ($z$ $<$ 1.5), include between 5 and 18 galaxies, and have line-of-sight velocity dispersions ranging from 168\,km\,s$^{-1}$ to 833\,km\,s$^{-1}$, with one double-peaked potentially merging structure identified at $z$ = 1.87.  
    \item Rest-frame colors, stellar masses, and star formation rates, demonstrate that the majority of member galaxies are actively star-forming and lie along the star-forming main sequence. 
    \item Detailed analysis of rest-frame U-band clumps reveals that 61.8$\%$ of our galaxies have off-axis clumps, with the clumpiest galaxies having  30 - 50$\%$ of their total flux concentrated in clumps.
    \item Comparing the global galaxy properties with the fraction of luminosity in clumps yields statistically significant positive correlations: more actively star-forming galaxies exhibit a higher fraction of their luminosity in clumps, which points to clumps being the dominant driver of star formation and stellar mass growth among our sample. 
\end{itemize}

Among our 11 structures, we found that the majority of galaxies are clumpy and highly star-forming, with no clear sign of evolution in the fraction of flux in clumps as a function of redshift.  Our strong statistical correlation between the fraction of rest-frame U-band luminosity in clumps and the SFR of the host galaxy highlights that clumps drive stellar mass growth at 1$ < z < $3.1, but particularly at $z >$ 2.25.  However, given that similar results have been found in samples of galaxies without environmental constraints, this seems to point toward similar clump formation mechanisms across environment, or similarities in the output populations.  

To continue exploring the role of clumpy protocluster galaxies at cosmic noon, we plan to take further advantage of the high resolution multi-band JWST photometry to perform simultaneous SED fitting of individual clumps.  This will yield robust constraints on clump ages, stellar masses, and star formation histories relative to their host galaxies, better constraining how dominant these clumps are in driving star formation activity.  Additionally, probing longer wavebands will allow us to identify both quiescent clumpy structures and dust-obscured star-forming clumps to more fully understand galaxy formation.  Furthermore, we will explore the build-up of proto-bulge clumps within this population and how they differ from the other identified star-forming clumps as this could be an additionally indicator of the evolutionary state of the cluster/protocluster galaxies.  Ultimately, a full treatment of clump formation and the role of environment requires explicitly comparing protocluster and field galaxies across redshift, mass, and SFR.  As such, we are planning a parallel exploration of how clumpy properties differ in the protocluster and field, which is vital to understand differences between the formation mechanisms of clumps (e.g., violent disk instabilities or mergers), particularly if protocluster galaxy clumps are primarily driven by galaxy-galaxy mergers.  However, probing the exact formation mechanisms of these galaxies requires high resolution IFU observations of both the galaxies and the surrounding medium are necessary to map the kinematics of these galaxies, allowing us to differentiate merger driven clump formation from gas driven instabilities. Beyond JWST, the horizons for galaxy clump science remain incredibly bright.  The ELT and the MICADO instrument \citep{Davies2016}, will have an angular resolution that is approximately five times smaller than JWST (0$\farcs$066 ; $\sim$ 560\,pc at $z$ = 2 in F200W).  As shown in Figure~\ref{Fig:Clumps-Radius}, the radius of the K-band PSF of MICADO will be $\sim$ 0$\farcs$012 ($\sim$ 100\,pc at $z$ = 2) \citep{Davies2016}, meaning the majority of detected clumps will be resolved.  Furthermore, the largest resolved clumps will now be broken down to sub-components, allowing us to further probe these building blocks \citep{Simioni2026}, which will lead to better constraints on how star-forming galaxies are building their stellar mass on the smallest scales.  

\begin{acknowledgements}
E.G.M. would like to thank Jesse Golden-Marx for useful discussions regarding statistical estimates and reading early drafts of this paper.  E.G.M. would also like to thank the organizers of the Tracing Cosmic Evolution with Galaxy Clusters V Conference, GalScale 2026 Conference, Cluster5 Conference, and the 3rd Edition Austrian ELT workshop for fostering stimulating environments that led to insightful discussions in shaping this paper.  E.G.M. would also like to thank Emiliano Merlin for useful discussions regarding JWST photometric analysis.\\
\\
PJW and BV acknowledge support from the INAF Mini Grant ``1.05.24.07.01 RSN1: Spatially-Resolved Near-IR Emission of Intermediate-Redshift Jellyfish Galaxies'' (PI Watson). 
PJW acknowledges support from the Agence Nationale de la Recherche (ANR), grant ANR-25-CE31-0192-01 (CIGALE-2D).\\
\\
B.C.L is supported by the international Gemini Observatory, a program of NSF NOIRLab, which is managed by the Association of Universities for Research in Astronomy (AURA) under a cooperative agreement with the U.S. National Science Foundation, on behalf of the Gemini partnership of Argentina, Brazil, Canada, Chile, the Republic of Korea, and the United States of America

\end{acknowledgements}
\bibliography{references}

\begin{appendix}
    \section{Spectroscopic Completeness}\label{Appendix}
\begin{figure}
    \begin{center}

    \includegraphics[scale=0.58,trim={0.35in 0.2in 0.0in 0.5in},clip=true]{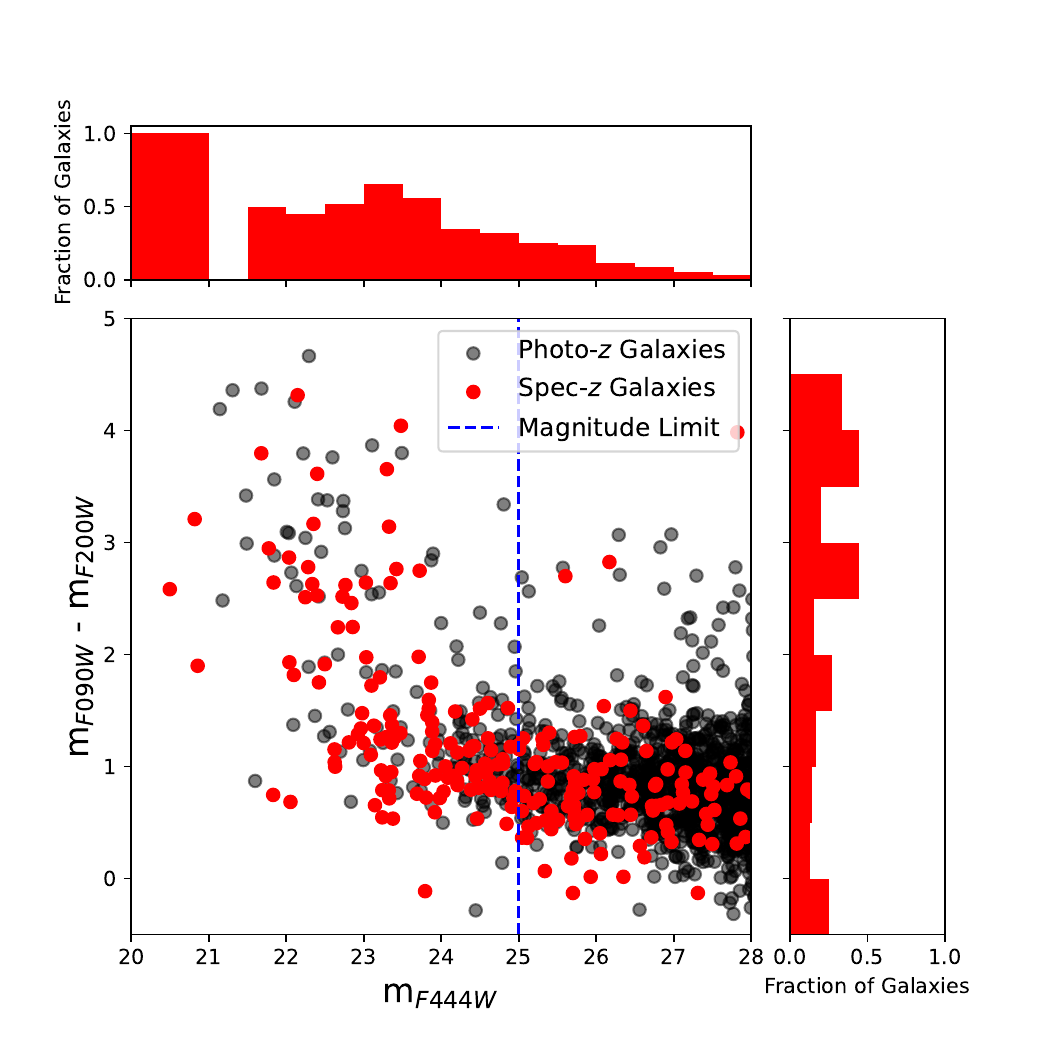} 
    \includegraphics[scale=0.58,trim={0.35in 0.2in 0.0in 0.5in},clip=true]{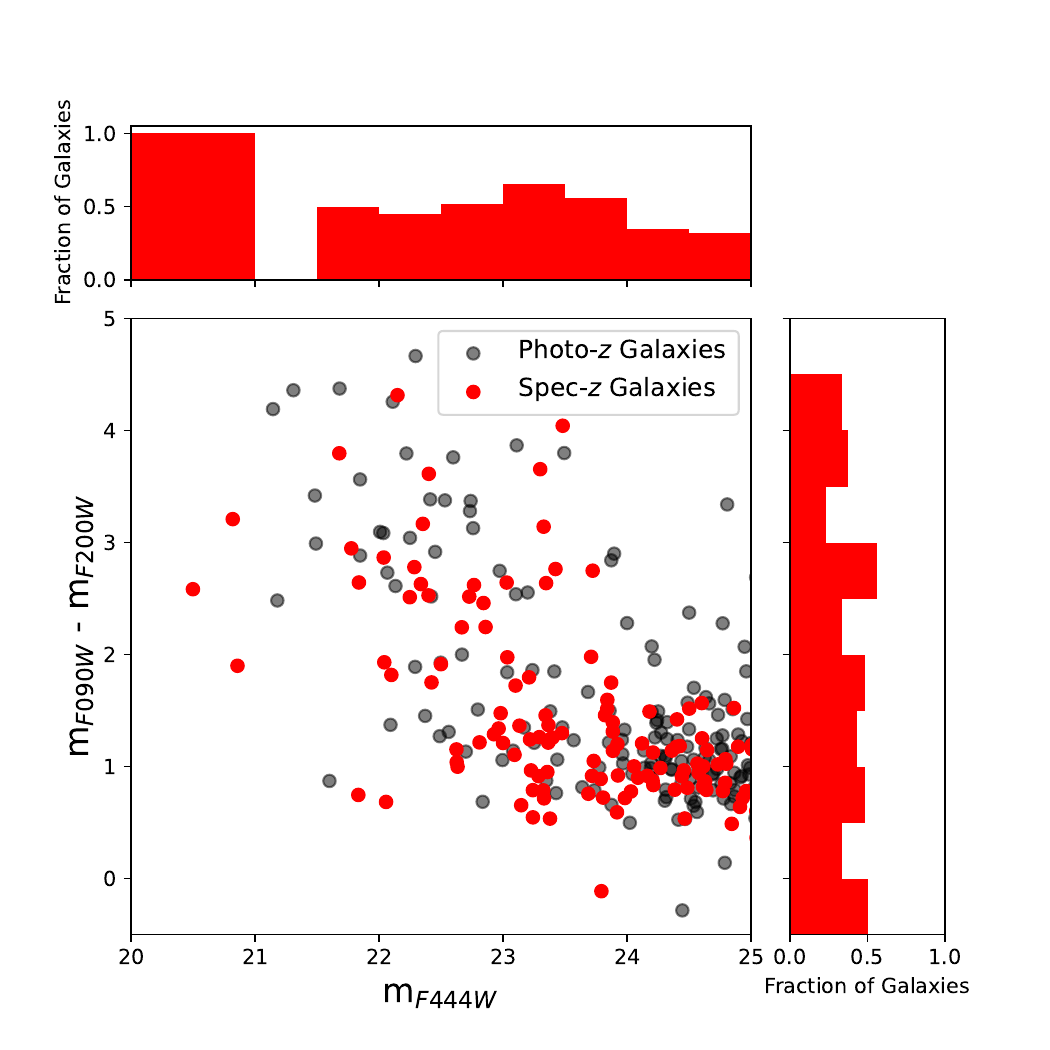}

    \caption{Color of spectroscopic and photometric galaxies as a function of F444W magnitude. Top: Magnification corrected colors and magnitudes for all galaxies at 2.2 $< z <$ 2.7 with both spectroscopic (red) and photometric (black) redshifts.  The histograms show the fraction of all galaxies that are have spectroscopic redshifts. The dashed blue line shows our magnitude limit.  Bottom: The same as the above plot, but limited to galaxies brighter than our magnitude limit, and thus reflecting any biases in our sample. }

\label{Fig:MagLim-CHECK}
\end{center}
\end{figure}

    As shown in Section~\ref{sect:data}, we limited our spectroscopic sample based on the photometric completeness of the spectroscopic galaxies.  This sample is likely biased against quiescent galaxies, given the requirements of our secure spectroscopy (see Section~\ref{sect:quiescent}.  To further explore these biases, we show, as an example, the sample of photometric and spectroscopic galaxies at 2.2 $< z <$ 2.7.  As can be seen in the top panel of Figure~\ref{Fig:MagLim-CHECK}, the fraction of galaxies with spec-$z$'s falls below 25$\%$ after our magnitude limit, where despite having a number of spec-$z$ galaxies, we no longer have a representative sample.  This trend highlights the necessity of requiring our magnitude limit for our analysis, particularly in regards to identifying potential structures.
    
    Importantly, we can use the y-axis histogram to probe the fraction of red, potentially quiescent, galaxies that have spectra.  As shown in the bottom panel of Figure~\ref{Fig:MagLim-CHECK}, where we only include the sample used in this work brighter than our magnitude limit, we find that among the reddest galaxies (m$_{F090W}$ - m$_{F200W}$ $>$ 2.0), between 20 and 50$\%$ of these galaxies have a spectra.  As most bins have less than the $\sim$ 50$\%$ completeness that we see among the bluer galaxy population, this is further evidence that the paucity of quiescent galaxies discussed in Section~\ref{sect:quiescent} may be due to the spectroscopic sampling, rather than these clusters intrinsically not hosting quiescent galaxies.    
    
\end{appendix}

\end{document}